\documentclass[a4paper,11pt]{article}
\usepackage{amssymb}
\usepackage{amsmath,amsfonts}
\usepackage{amsthm}
\usepackage[mathscr]{eucal}
\usepackage{graphicx}
\usepackage{epsfig}
\usepackage{amsbsy}
\usepackage[bf, small]{caption}

\newfont{\AMSL}{msbm10 scaled \magstep 1}

 \newtheorem{Theorem}{Theorem}[section]

 \theoremstyle{Definition}
 \newtheorem{Definition}{Definition}[section]
 \numberwithin{equation}{subsection}
 \newtheorem{remark}{Remark}

\begin{document}

\date{}
\title{ Kohonen neural networks and genetic classification}
\author{ Daniela Bianchi \thanks{
Department of Physics, University "La Sapienza", Rome; E-mail:
danielabianchi12@gmail.com}\and Raffaele Calogero\thanks{%
Bioinformatics and Genomic Unit, University of Turin, Turin;
E-mail:
raffaele.calogero{@}unito.it} \and Brunello Tirozzi\thanks{%
Department of Physics, University "La Sapienza", Rome; E-mail:
brunello.tirozzi{@}roma1.infn.it} } \maketitle

\abstract{We discuss the property of a.e. and in mean convergence
of the Kohonen algorithm considered as a stochastic process. The
various conditions ensuring the a.e. convergence are described and
the connection with the rate decay of the learning parameter is
analyzed. The rate of convergence is discussed for different
choices of learning parameters. We proof rigorously that the rate
of decay of the learning parameter which is most used in the
applications is a sufficient condition for a.e. convergence and we
check it numerically. The aim of the paper is also to clarify the
state of the art on the convergence property of the algorithm in
view of the growing number of applications of the Kohonen neural
networks. We apply our theorem and considerations to the case of
genetic classification which is a rapidly developing field.}

\section{Introduction}

Data clustering (\cite{[1]}-\cite{[5]}) is a basic technique in
gene expression data analysis since the detection of groups of
genes that manifest similar expression patterns might give
relevant information. Therefore it is important to have a good
control on the properties of clustering algorithms. The Kohonen
algorithm ( or Kohonen neural network)(\cite{[6]}-\cite{[8]}) is
currently used in this field. The Kohonen neural networks are
different from the other neural networks like back propagation or
the Hopfield model (\cite{[9]}-\cite{[12]} ). The main difference
is that there is only a single layer of units ( named neurons) and
the output of the network is just a vector or a scalar associated
with each neuron called weight vector. These networks are commonly
used for classifying sets of experimental data. The weight vector
associated with the neuron represents a characteristic vector of a
certain subset of the data. The set of these subsets constitutes a
disjoint partition of the measures. The sets of the partition are
also called clusters and in the applied science there are many
different algorithms which construct clusters from a data set.
Many of these algorithms have the drawback that they depend on
arbitrary choice of some parameters and therefore the clustering
results might be non unique. The main feature of clustering by
means of the Kohonen algorithm is that it depends only on the
choice of a special function, the $ ${\it learning parameter},
which has been extensively characterized. The process of
individuation of the weights is called the $ ${\it learning
process} and the Kohonen algorithm is a special learning process.
This algorithm consists in extracting at each time step $n$ a
number or a vector from the data set and subsequently the nearest
weight to this data  is modified of a quantity proportional to the
difference among these two vectors multiplied by a parameter. This
parameter is the learning parameter, and it must decrease with
$n$. The convergence of the learning process strongly depends on
the rate of decay of the learning parameter and the investigation
of this point is one of the main topics of this paper.  An
important characteristic of the Kohonen algorithm is the {\it Self
Organization}(SO) which can be understood as the fact that the
sequence of the weights converges to a unique limit independently
from the chosen sequence of the data presented to the network and
from the initial values of the weights. In the language of the
stochastic processes we can express this fact by observing that
the sequence of the weights is a stochastic process and SO is
equivalent to the a.e. convergence of the learning process. This
property is rather strong and it is supposed to hold in many
applications of the Kohonen networks but unfortunately it is not
trivial at all. This is especially true for genetic application
where the set of clusters (atoms) describes different cell
conditions or different genes function. In order to have a real
biological meaning the classification should be independent on the
initial conditions of the weights and from the input sequence. So
it is worth investigating, both theoretically and numerically, the
connection among the a.e. convergence and the possible choices of
the learning parameter $\eta(n)$ and the different versions of the
Kohonen algorithm. There are already many important results on
this subject ([13]-[28]). All these results show that there is a
critical dependence of the a.e. convergence on the probability
distribution of the data, on the choice of the learning algorithm
and on the velocity to approach zero by $\eta(n)$ . In this paper
we generalize the results obtained in the paper of Feng and
Tirozzi (\cite{[25]}) relaxing the condition on the convergence of
the series $\sum_n \eta(n)^2$ ( but of course it is assumed that
$\eta(n) \rightarrow 0$), so only the condition $\sum_n \eta(n)
\rightarrow \infty$ remains. The condition $\sum_n
\eta(n)^2<\infty$ is used in all the other versions of the
theorems of convergence, but we have verified numerically that it
implies a too rapid convergence to zero of the learning parameter.
So the good decrease rate for $\eta(n)$ is to go to zero more
slowly than $ 1/n$.But our theorem does not exclude the $1/n$
decay rate since it also satisfies the condition $\sum_n \eta(n)
\rightarrow \infty$. Thus this theorem gives a support to the
property of a.e. convergence for the right decay of $\eta$ but it
is uncompleted  because we cannot show that the stronger condition
$\sum_n \eta(n)^2 < \infty$ spoils the a.e. convergence. The
previous results  also are troublesome because we are faced with
the fact that a theorem with a defined proof of convergence does
not correspond to the numerical simulations. The only thing we can
say is that at least our version of the convergence theorem picks
the right decrease property. There is a well known general
explanation about the right choice of the rapidity of the learning
parameter decay which is connected with the existence of
meta-stable points. In analogy with the Simulated Annealing (SA)
we can say that the learning parameter corresponds to the
temperature and it is a well known fact that a too rapid decrease
of the temperature in the SA makes the algorithm stop on the local
minima of the energy function. The unlucky situation is that in
the case of the Kohonen algorithm there is no such a function. In
many proofs of the convergence one can find some functional with a
similar property but they are not the energy or the Liapunov
function. The other important question tackled in this paper is
about the rate of the convergence of the algorithm: since the
condition $\sum_n \eta(n) \rightarrow \infty$ can be satisfied by
many different $\eta(n)$ we compare the different choices
analyzing the velocity of approach to the limit of the
corresponding algorithm. This question is important in any case
but has special relevance in gene clustering where the data set is
the set of expression levels of $M$ genes, $M$ being rather large
thank to the application of the microarrays technique. The meaning
of $M$ in our construction is the maximum number of iterations of
the learning process. Another question considered in this paper is
the analysis of the relation of the a.e. convergence with the
probability distribution of the data and also with the different
versions of the Kohonen algorithm. We then apply all these results
to the problem of clustering and classifying the great number of
genes revealed in the microarrays experiments. The possibility of
applying clustering algorithms ( not only the Kohonen algorithm)
in genetics appeared with the development of the DNA microarrays
technology. The micro-array allows to monitor simultaneously the
expression levels of thousands of genes during important
biological processes. Elucidating the patterns hidden in the gene
expression data is a tremendous opportunity for functional
genomic. However, because of the large number of genes and
complexity of biological networks, it is difficult to interpret
the resulting mass of data; so the clustering techniques become
essential in data mining process for identifying interesting
distributions and patterns in the underlying data.\\
Clustering algorithms have simplified the grouping of genes with
similar biological expression. Co-expressed genes found in the
same cluster suggest functional similarities. Gene clustering also
becomes the first step to uncover the regulatory elements in
transcriptional regulatory networks. Co-expressed genes in the
same cluster might be involved in the related cellular process and
strong expression pattern correlation between those genes
suggests co-regulation.\\
There is a large literature on cluster analysis and genetic one
(\cite{[1]} - \cite{[5]}); numerous approaches were proposed on
the basis of different quality criteria and not all the algorithms
are well founded. In addition the results of the algorithms depend
strongly on many arbitrary choices, for example on the initial
conditions and the value of
the threshold.\\
The main topic of the last section of this paper is an application
of the Kohonen algorithm to a concrete problem of gene
classification. The aim is to find the genes which are over
expressed during the treatment of tumor cells of mice using a
clustering technique that has the
minimum arbitrary choices. \\
The analysis made in the first sections of this paper convinced us
to use  the Kohonen algorithm.\\
We compare the results obtained with the Kohonen algorithm to this
problem with the ones obtained using the PCA (Principal component
analysis) and Hierarchical clustering algorithm (\cite{[30]}).
This is the first step of a larger work of comparing the results
of gene classification obtained by means of different algorithms.
We think that this work is necessary in
order to validate the gene clustering.\\
Another important issue is the variability of the expression
levels of genes obtained by different samples which cannot be
considered equal. For economic and time reasons it is difficult to
have more than three biological realizations of the experiment and
this is the origin of an error in the data. The errors influence
the structure of the clusters so it is possible that a gene
changes cluster if we take into account this error in the
analysis. In our work we have included explicitly this effect and
evaluated its influence on the
results.\\
The structure of the paper is the following. In Section $2$, after
a short intuitive introduction, we show the algorithm and explain
its properties using a precise mathematical formulation, enunciate
the theorems and give the proofs . In Section 3 we show the
results of the numerical simulations. In Section 4 we show the
applications to the mice data and our results. In Section 5 we
give our conclusions.

\section{The Kohonen Network}

\subsection{An intuitive description }

The Kohonen Network (\cite{[6]}-\cite{[8]}) is formed by a single
layered neural network. The data are presented to the input and
the output neurons are organized with a simple neighborhood
structure. Each neuron is associated with a reference vector ( the
weight vector), and each data point is "mapped" to the neuron with
the "closest" ( in the sense of the Euclidean distance) reference
vector. In the process of running the algorithm, each data point
acts as a training sample which directs the movement of the
reference vectors towards the value of the data of this sample.
The vectors associated with neurons, called weights, change during
the learning process and tend to the characteristic values of the
distribution of the input data. The characteristic value of one
cluster can be intuitively understood as the typical value of the
data in the cluster and will be defined more precisely in the next
subsections. At the end of the process the set of input data is
partitioned in disjoint sets ( the clusters) and the weight
associated with each neuron is  the characteristic value of the
cluster associated with the neuron in one dimensional case, which
is the case of interest to us. We limit our analysis to this case
because the condition of convergence of the algorithm is easier to
check, the cluster of the partitions are easier to visualize and
it is not difficult to compare the behavior of the genes in the
clusters corresponding to the different biological conditions.
Each neuron individuates one cluster, the physical or biological
entities with measure values belonging to the same cluster are
considered to be involved in the same cellular process. Thus the
genes with expressions belonging
to the same cluster might be functionally related.\\
The following points show the main properties which make the
Kohonen network useful for clustering :
 \begin{enumerate}

    \item Low dimension of the network and its simple
    structure.
    \item Simple representation of clusters by means of vectors associated with each neuron.
    \item Topology of the input data set is somehow mapped in the topology of
           the weights of the network.
    \item Learning is unsupervised.
    \item Self-organized property

 \end{enumerate}

The points 1)-2) are simple to understood and many examples are
shown in the Section 3. The point 3) means that neighboring
neurons have weight vectors not very different from each other.
The point 4) means that there is no need to have an external
constraint to drive the weights towards their right values beside
the input to the network and that the learning process finds by
itself the right topology and the right values. This holds only if
the learning process with which the network is constructed
converges a.e. or if the mean values are taken. The
self-organization is formulated in the current literature
referring to some universality of the structure of the network for
a given data set. It is connected to the point 3) and is also a
consequence of a.e. convergence or of the convergence of the
average of the weights over many different learning processes.

\subsection{Exact definition}

 In this subsection we give the definitions using exact
 mathematical terms. We restrict ourselves to the particularly
 simple one dimensional case which is the most interesting for our
 applications.
 First we show how the Kohonen network is used for classification
 and then what is the process of its construction.
 Let $I= I_1,….,I_N$  be a partition of the interval $(0,A)$ of the
 possible values of the expression levels in the intervals $I_i$,
 $ \bigcup_i I_i= (0,A)$  and $I_i \bigcap I_j = 0 $.\\
 Suppose that the construction of the Kohonen network has been already
 done and the $I_i$ are the clusters.\\
 Let $\omega_i, \quad i=1,…N$ be the weights or the characteristic values
 of the clusters which will be exactly defined below.
 Then a data $\xi$ is said to have the property $i$ if $\xi \in I_i$.\\
 The classification error  is
\begin{eqnarray*}
      |\xi-\omega_i|
\end{eqnarray*}
Then the global classification error E of the network is
\begin{equation}\label{err}
    E=\frac{1}{T}\sum_{i=1}^N\int_{I_i} \|\xi-\omega_i\|^2 f(\xi)
    \,d\xi
  \end{equation}
where $f(\xi)$ is the density of probability distribution of the
input data and $T=\sum_{i=1}^N|I_i|$,with $|I_i|$ is the number of
data in the set $I_i$.\\
The partition $I= I_1,….,I_N$ is optimal if the associated
classification error $E$ is minimal. The characteristic vectors
$\omega_i$ are the values which minimize $E$. Before giving exact
definitions let us explain in simple terms the procedure for
determining the sets $I_i$ and the associated weights $\omega_i$.
Let $x(1),\dots,x(P)$ be a sequence of values randomly extracted
from the data set, distributed with the density $f(x)$ and take
randomly the initial values $\omega_1(0),...,\omega_N(0)$ of the
weights. When an input pattern $x(n)$, $n=1,..,P$, is presented to
the network all the differences $|x(n)-\omega_i(n-1)|$ are
computed and the winner neuron is the neuron $j$ with minimal
difference $|\xi(n)-\omega_j(n-1)|$. The weight of this neuron is
changed in a way defined below, or, in some cases, the weights of
the neighboring neurons are changed. Then this procedure is
repeated with another input pattern $x(n+1)$ and with the new
weights $\omega_i(n)$ until the weights $\omega_i(n)$
converge to some fixed values for $P$ large enough.\\
In this way we get a random sequence
$\omega(n)=(\omega_1(n),...,\omega_N(n))$ which converges a.e,
under suitable conditions on the data set, with respect to the
choice of the random sequence of data and the random choice of the
initial conditions of the weights, $n$ is the number of iterations
of the procedure. The learning process is the sequence $\omega(n)$
and the S.O. (self-organizing property) coincides in practice with
the almost everywhere convergence of $\omega(n)$. The learning
process converges somehow to the optimal partition in the Kohonen
algorithm. In fact the algorithm can, with some approximation, be
viewed as a gradient method applied to the function $E$:\\
\begin{eqnarray*}
  \omega_i(n+1)
  =\omega_i(n)+\eta(n)[\xi(n)-\omega_i(n)]\\
  \sim \omega_i(n)-\frac{1}{2}\eta(n)\nabla_{\omega_i(n)}
  \mathrm{E}\\
\end{eqnarray*}
In one dimension the Kohonen algorithm in the simplest version of
the  winner-take-all case is :
\begin{enumerate}
    \item Fix $N$.
    \item Choose randomly at the initial step ($n=0$) the $\omega_i$ ( $0\leq i \leq N$).
    \item Extract randomly the data $\xi(1)$ from the data set.
    \item Compute the modules
          \begin{eqnarray*}
                    |\omega_i(0)-\xi(1)| \;\;\; i=1,\ldots,N
           \end{eqnarray*}

    \item Choose the neuron $v$ such that
           \begin{eqnarray*}
                   |\omega_v(0)-\xi(1)|
           \end{eqnarray*}
    is  the minimum distance. $v $ is the winner neuron

    \item Update only the weight of the winner neuron:
          \begin{eqnarray*}
       \omega_v(1) = \omega_v (0) + \eta(1) (\xi(1) -  \omega_v(0) )
          \end{eqnarray*}
    \item $n =n+1$

\end{enumerate}
One of the basic property of the Kohonen network is that the
weights are ordered if the learning process converges. \\
We remind the definition of the order of a one dimensional
configuration:
\[
|r-s|<|r-q|
\Leftrightarrow|\omega_r-\omega_s|<|\omega_r-\omega_q|,\ \forall
r,s,q\in\{1,2...,N\}
\]
the order holds also for the other inequality
$|\omega_r-\omega_s|>|\omega_r-\omega_q|$. Then:\\
\texttt{The ordering property is}:\\ \\
\textit{If the Kohonen learning algorithm applied to one
dimensional configuration of weights converges the configuration
order itself at a certain step of the process. The same order is
preserved at each subsequent step of the
algorithm}\\ \\
This property allows to check when the algorithm converges since
the final configuration of weights must be also ordered and it is
a necessary property for a.e. convergence. We also check the
remarkable property proved by Kohonen (\cite{[6]}-\cite{[8]}) that
the mean process $\omega(n)$, i.e. the process obtained by
averaging with respect different choices of the sequence
$x(1),...x(P)$ is always converging. But for getting the a.e.
convergence from the convergence of the mean values additional
hypothesis must be used and the discussion and the applications of
these results to a case of genetic classification is the main
topic of this paper.

\subsection{General formulation }

\begin{figure}[!h]
\begin{center}
\includegraphics[height=5cm,width=7cm]{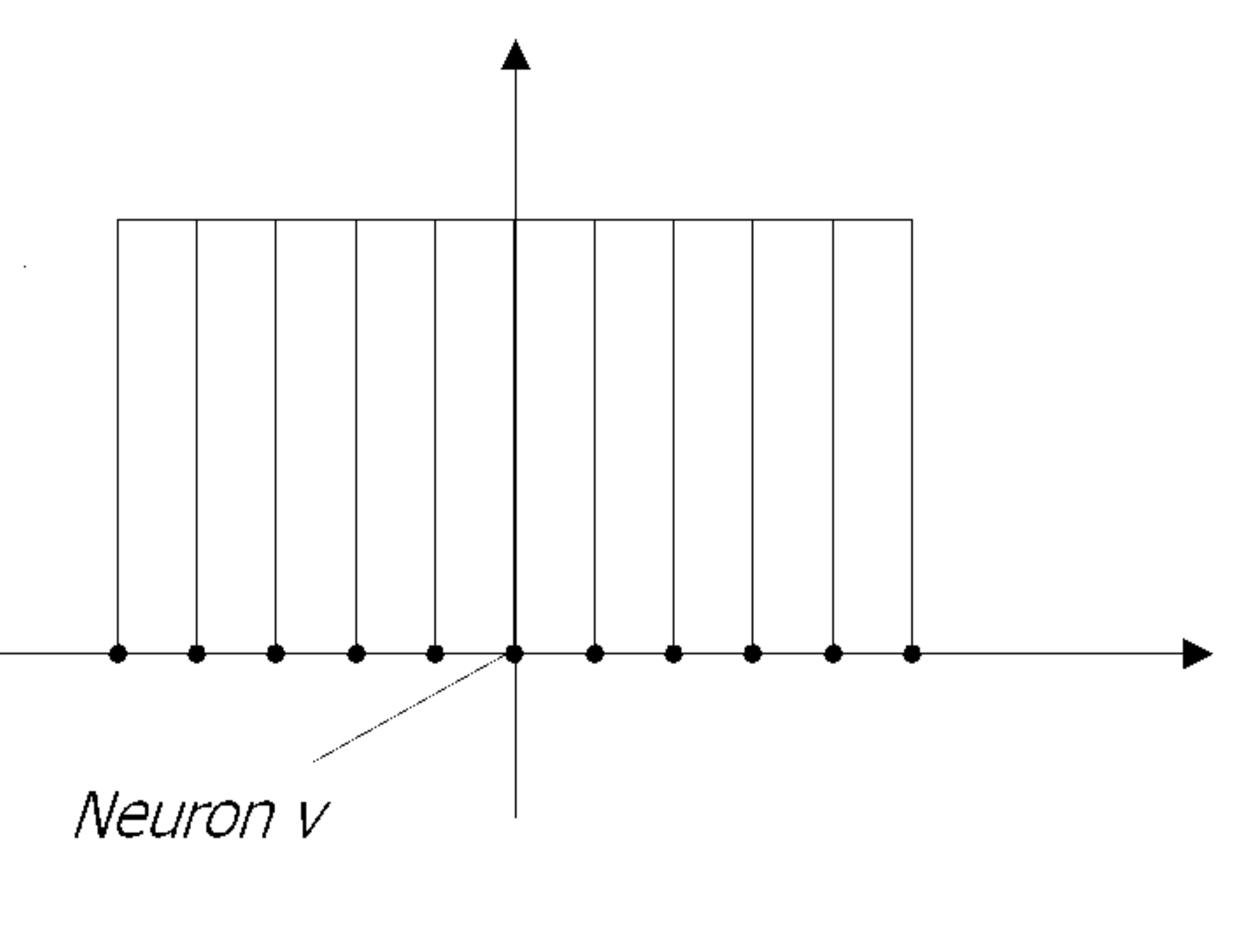}
\caption{Neighborhood function $\Lambda(i,v)$} \label{lambda}
\end{center}
\end{figure}

\begin{figure}[!h]
\begin{center}
\includegraphics[height=5cm,width=7cm]{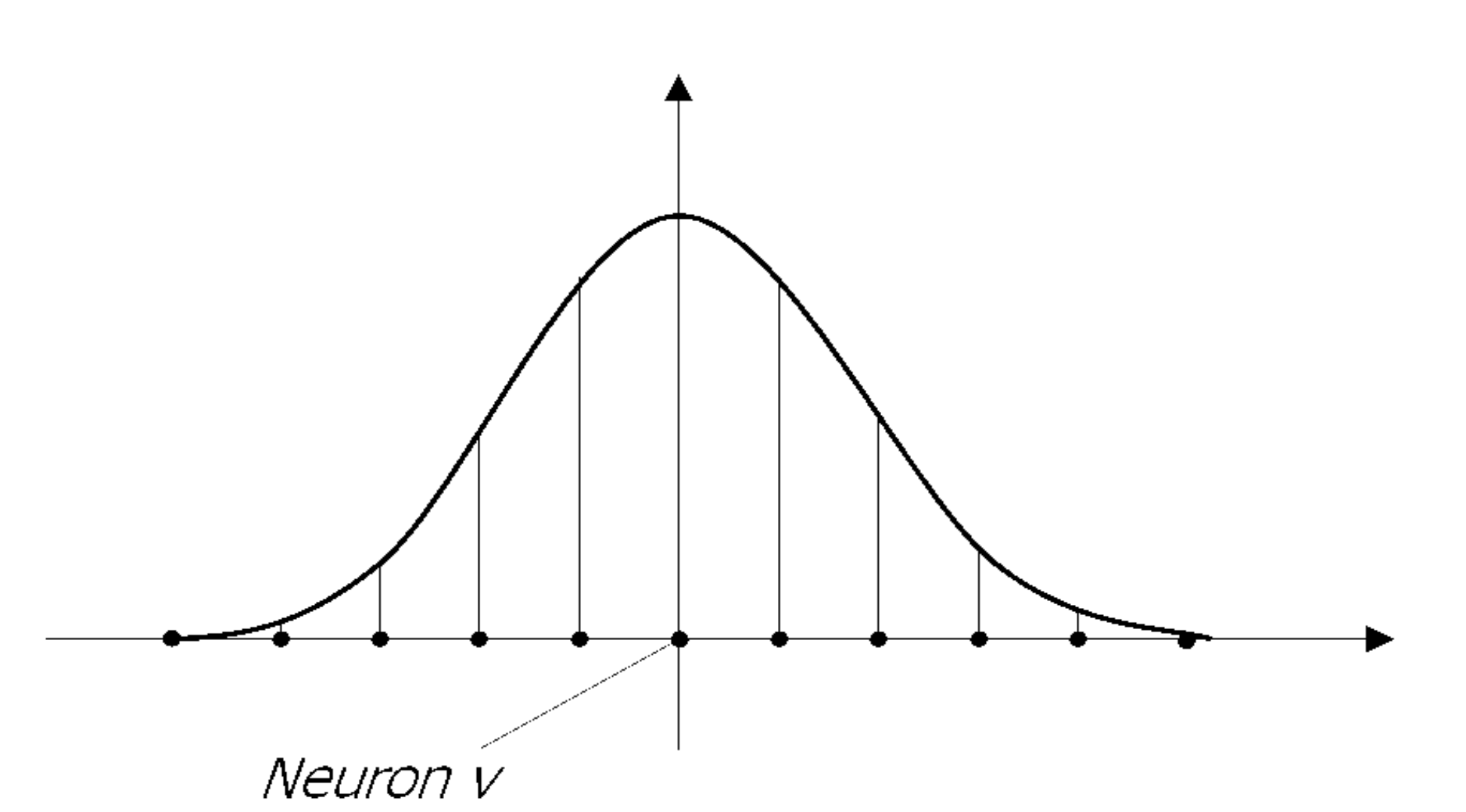}
\caption{Neighborhood function $ h(i,v,n)$} \label{acca}
\end{center}
\end{figure}

We describe now the Kohonen algorithm in more general terms for
allowing the treatment of all the possible cases.\\
The Kohonen network is composed by a single layer of output units
$\mathcal{O}_i$, $i= 1,...N$ each being fully connected to a set
of inputs $\xi_j(n)$, $ j=1,...,M$. An M dimensional weight vector
$\omega_i(n)$,$\omega_i(n)=(\omega_{ij}(n),j=1,...M)$ is
associated with each neuron.  $n$ indicates the $n$-th step of the
algorithm.\\
We assume that the inputs $\xi_j(n)$,$ j=1,...,M$ are
independently chosen according to a probability distribution
$f(x)$. For each input $\xi_j(n)$,$ j=1,...,M$ we choose one of
the output units, called the \emph{winner}. The \emph{winner} is
the output unit with the smallest distance between its weight
vector $\omega_v(n)$ and the input
\[
     ||\omega_v(n-1)-\xi(n)||
\]
where $||.||$ represents Euclidean norm. Let
$\mathrm{\overline{I}}(.,.)$ be the function
\[
     \mathrm{\overline{I}}(\omega_v(n),\xi(n+1))=\mathrm{I_{\{||\omega_v(n)-\xi(n+1)||<
     ||\omega_j(n)-\xi(n+1)||,\ j \neq v \}}}
\]
where $\mathrm{I_A}$ is the characteristic function of the event
$A$, i.e, $\mathrm{I_A}(x)=1$ if $ x\in\mathrm{A}$ and
$\mathrm{I_A}(x)=0$
if $x\not\in\mathrm{A}$.\\
This function selects the event in which  the weight of the neuron
$v$ is the nearest to the input data $\xi(n)$ and it is necessary
for writing the learning process in a compact form. The
generalized Kohonen algorithm updates the weights of the neurons
belonging to a given neighbor of the \emph{winner} neuron:
\begin{equation}\label{2.1}
\omega_{ij}(i+1)=\omega_{ij}(n)+\eta(n)\Gamma(i,v)\mathrm{\overline{I}}(\omega_v(n),\xi(n+1))\
  (\xi_j(n+1)-\omega_{ij}(n))
\end{equation}
 $i=1,....,N$ e $j=1,...M$ or in vector form
\begin{equation}\label{2.2}
  \omega_i(n+1)=\omega_i(n)+\eta(n)\Gamma(i,v)\mathrm{\overline{I}}(\omega_v(n),\xi(n+1))\
  (\xi(n+1)-\omega_i(n))
\end{equation}
where $ \eta(n)$ is the positive learning parameter $\eta(0)<1$,
$\eta(n)\geq\eta(n+1)$ and $\Gamma(i,v)$ is a non increasing
function of $|i-v|$, the distance among the neuron $i$ and $v$ on the
lattice where the neurons of the network are located.\\
This version is more general than the winner-take-all rule
explained before. Not only the weight of the winner neuron is
updated but also the weights of the neurons which belong to a
neighborhood defined by the function $\Gamma(i,v)$. We discuss
various choices of the function $\Gamma(i,v)$ below. After the
learning procedure is finished, the set of input vector will be
partitioned into non overlapping clusters. This means that a new
signal $\xi(n+1)$ is classified as the pattern $i$ if and only if
\[
      ||\omega_i-\xi(n+1)||\leq||\omega_j-\xi(n+1)||,\  j\neq i
\]
Let us introduce the definition of Voronoi tessellation
$\Pi(y)=(\Pi(y)_i,i=1,...N )$  associated with a family vectors
$y_1,...,y_N \in\Omega$ , $\Omega$ being a given compact of
$\mathbb{R}^M$ .
\begin{Definition}
For a given compact subset $\Omega \in \mathbb{R}^M $, the Voronoi
tessellation $\Pi(y)=(\; \Pi(y)_i$, $\;\; i=1,..N)$ associated
with a family of vectors $y_1,....,y_N$ is the partition of
$\Omega$:

\begin{equation}\label{tassel}
  \Pi(y)_i=\{x,\|y_i-x\|\leq \|y_j-x\|,j\neq i \} \;\;\; i=1,...N
\end{equation}
\end{Definition}Therefore a Voronoi cell of an unit $i$ contains those vectors
which are closer to the weight $\omega_i$ than to the other
weights. The characteristic values  mentioned before are the limit
of the sequences of the vectors $\omega_i(n)$ defined by the above
algorithm and are weights of the Voronoi tessellation obtained in
the limit.\\
A crucial point of the algorithm is the choice of the neighborhood
function $\Gamma(i,v)$ of the \emph{winner} neuron. It determines
the region around the \emph{winner} neuron where there are the
neurons which update their weight vectors together with the winner
neuron. A convenient choice is the finite region of activation of
the winner neuron, i.e. $\Gamma = \Lambda$ where :
$$
  \Lambda(i,v)=
 \{\begin{array}{c}
   1 \ if \ | i-v| \leq s \;\;\;\; \\
   0 \ otherwise \;\;\;\; \\
  \end{array} \label{2.3}
$$
where $|.|$ represents the distance between the neuron $i$ and
the \emph{winner } neuron $v$.\\
If $s=1$ and the neural network is one dimensional, the region of
activation includes the \emph{winner} and the two nearest units
(figure \ref{lambda}); if the network is designed in two dimension
then the range includes the eight nearest neighbor units near the
\emph{winner} .\\
If $\Lambda(i,v)=\delta_{iv}$ the algorithm coincides with the
\emph{winner takes all} algorithm we described in the previous section.\\
\\
Another choice, often used in the applied research, for the
neighborhood function $\Gamma$ is a gaussian function $h$ defining
a region around the winner neuron with amplitude decreasing with
the number of iterations of the learning process:
\begin{equation}\label{2.4}
h(i,v,n)=\exp(- \frac{|i-v|^2}{\sigma(n)^2})
 \end{equation}
where $\sigma(n)$ is a decreasing function. A commonly used choice
is:
\[
\sigma(n)=\sigma_i (\frac{\sigma_f}{\sigma_i})^\frac{n}{n_{max}}
\]
where $n_{max}$ is maximum number of iterations of the algorithm
and  $\sigma_f$, $\sigma_i$ are respectively the final and initial
value of the parameter $\sigma$(figure \ref{acca}).

\subsubsection{ The theorem of convergence}

The first result about the algorithm convergence was found by
Kohonen (\cite{[7]}). He concentrated on one-dimensional mapping
and demonstrated  that the weights converge in mean to the limit
values. Although the result is enunciated as a.e. convergence in
this paper only the convergence in mean is proven. The convergence
in mean is obtained by making the average of the weights on many
different sequences of patterns $x(n)$. The ordering of the
weights has been
proved  in (\cite{[7]}) for the winner-take-all process. \\
In the paper of Erwin et al. (\cite{[16]},\cite{[17]}) there is a
proof of ordering for one-dimensional case which holds for any
neighborhood function which is monotonically decreasing with
distance and in the case of non uniformly distributed input.\\
Many other authors (\cite{[13]},\cite{[19]},
\cite{[21]},\cite{[22]}, \cite{[23]}, \cite{[25]}, \cite{[28]},)
investigated the convergence properties of the Kohonen algorithm
in one and more dimensions, someone by viewing the weight values
as states of a Markov process, others using the ordinary
differential equations for the mean values of the network. But the
main results have been limited to one dimensional map where the
property of order is valid and under certain conditions on
$\eta(n)$, the expectation of the values weights converges to a
unique value. The existence and uniqueness of the minimum is
ensured by the existence of a unique minimum of some functional,
but the existence of the minimum is difficult to check for non
uniform distribution of the
input values especially in the multidimensional case.\\
In more than one dimension, despite the robustness of the
algorithm which has been used successfully in many different
application area, there is still no proof of a necessary and
sufficient condition for the convergence of the algorithm. There
are proofs of sufficient conditions and only a few for the multi
dimensional case, see for example Feng and Tirozzi (\cite{[25]}),
Lin and Si (\cite{[19]}), Sadeghi (\cite{[23]}). Lin and Si have
shown that the distribution of the weight values converge to a
stationary state introducing and studying the same objective
function proposed by Ritter and Schulten \cite{[21]}. In the paper
of Feng and Tirozzi the convergence problem of the Kohonen feature
mapping algorithm has been proven by using stochastic
approximation theory.  But in all these papers the rate of
decrease of the learning parameter is too fast and so these
theorems are contradicted by numerical results. Only in the paper
of Feng and Tirozzi it is mentioned explicitly that the rate of
decrease of the learning parameter of these theorems is too fast
and there is a proposal for a slower decay. In this paper we proof
that there is a.e. convergence if the rate is the one of numerical
simulations, but we can show only the sufficiency of this
condition. Moreover a condition of the existence
of a global attractive minimum is always required.\\
If there is no global minimum there is no a.e. convergence and the
algorithm remains stacked, as in the case of simulated annealing,
in some points which might not even be ordered and then the
convergence is obtained only by averaging with respect to the
sequences of learning examples, which is happening for the genetic
data in general. Now we start to expose the definitions and
concepts used in our proof. We first explain the definitions used
in the book of Nevel'son and Has'minski (\cite{[20]}) which will
be used in the proof of our main theorem. Let $ \xi(k),\ k\leq n$
be the sequence of random patterns presented to the network during
the learning and $\mathcal{F}_n$ the $\sigma$-algebra generated by
them, $\mathbb{E}(\zeta|\mathcal{F}_n)$ is the conditional
expectation of the random variable $\zeta$
with respect to the sigma algebra $\mathcal{F}_n$.\\
Our aim is to prove that the process of the weights $\omega(n)$
converges to a certain set $B\subseteq\mathbb{R}^{N\times M}$ (the
limit set), so we need the definitions summarized in the following list:\\
\begin{Definition}.\\
\begin{enumerate}
    \item A distance between vectors $y \in \mathbb{R}^M$ and $\omega_i \in \mathbb{R}^M$ $\rho(\omega_i,y)$, with $i=1,...N$.
    \item A distance from the point $\omega(n)$ and the set $B$: $\rho(\omega(n),B)=inf_{y\in B} \rho(\omega,y)$.
    \item An $\varepsilon$ neighborhood of B, $U_\varepsilon(B)=\{\omega:\rho(\omega,B)<\varepsilon\}$.
    \item The complementary set of this neighborhood $V_\varepsilon(B)=\mathbb{R}^{N\times M}\backslash U_\varepsilon(B)$.
    \item The intersection of the complementary set with a sphere of radius R: $V_{\varepsilon,R}(B)=V_\varepsilon(B)\cap\{\omega(n):|\omega(n)|<R\}$.
    \item A positive definite Lyapunov function $ W(n,\omega)$, $\omega\in \mathbb{R}^{N\times M}$.
    \item An operator $LW(n,\omega)=\mathbb{E}\ (W(n+1,\omega(n+1))-W(n,\omega(n))|\mathcal{F}_n)$ defining a kind of first difference of the Lyapunov
    function by means of conditional expectation.
    \item A negative function $g(n,\omega)$ used for bounding the
    increments of the Lyapunov function $ W(n,\omega)$ such that
     \begin{equation}\label{condoper2}
      \inf_{n\geq Q, \omega\in V_{\varepsilon,R}(B)}[-g(n,\omega)]>0
     \end{equation}
     for all $R>\varepsilon>0$ \ and some $Q=Q(\varepsilon,R)$.
\end{enumerate}

\end{Definition}
Let us briefly comment these definitions. \\
1) As we have seen before $\omega_i$ and $y$ are $\mathbb{R}^{M}$
vectors and since we have to compare their difference
it is necessary to introduce the module of these vectors.\\
2) In the general case the limit point might be a set so the
distance of a point from a set must be defined.\\
3) As is usual in the theory of limits one needs to find a
neighborhood $U_\varepsilon(B)$ of the limit points $B$ which
differ from $B$ by a small portion.\\
4) It is also necessary to introduce the complementary set
$V_\varepsilon(B)$ of this
neighborhood.\\
5) For doing the estimates of asymptotic limits of series or
functions it is useful to introduce a spherical subset
$V(B)_{\varepsilon,R}$ of $V_\varepsilon(B)$.\\
6) In analogy with the theory of stability in order to show the
convergence of a trajectory of a dynamical system it is useful to
have a Liapunov function and compute its increments. In this case
we do not have an usual dynamical system but a stochastic
sequence.\\
7) The consequence of this fact is that the derivative ( or
increments) of the Liapunov function is not the usual one but is a
conditional expectation. The convergence holds for the sequence
$W(n, \omega)$ as a consequence of Doob's theorem of convergence
for martingales but it is difficult to use the concepts of local
and global minimum in this situation. In our theorem the concept
of global minimum in the classical sense is introduced but for the
bounding
function $g(n,\omega)$.\\
We will use this theorem of (\cite{[20]}) in our proof of the a.e.
convergence:\\
\begin{Theorem}\label{teoremaoper}
Suppose that there exist a function  $W(n,\omega) \geq 0$ such
that:
\begin{equation}\label{condoper1}
  LW(n,\omega)\leq \eta(n) g(n,\omega)
\end{equation}
where $n\geq 0$, $\ \omega\in \mathbb{R}^{N\times M}$ and $g$ the
function which satisfies the above statement (\ref{condoper2})\\
Moreover let :
\begin{equation}\label{condoper3}
  \sum_{n=1}^{+\infty}\eta(n)= +\infty \ , \; \eta(n)>0
\end{equation}
and:
\begin{equation}\label{condoper4}
  inf_{n\geq 0} W(n,\omega)\longrightarrow+\infty \, \;
  |\omega|\rightarrow+\infty
\end{equation}
Then, considering the previous definitions:
\begin{equation}\label{condoper5}
\mathrm{P}\{\sup_n|\omega(n)|=R<+\infty\}=1
\end{equation}
\begin{equation}\label{condoper6}
\mathrm{P}\{ \sum_{u=0}^{+\infty} \eta(u) [-g(u,\omega(u))]<
+\infty\}=1
\end{equation}
\begin{equation}\label{condoper7}
\mathrm{P}\{\liminf_{n\rightarrow+\infty} \rho(\omega(n),B)=0\}=1
\end{equation}

\end{Theorem}We can say that a random process $\omega(n)$ converges a.e. to a
limit set $B$ if it is possible to find a Liapunov function
$W(n,\omega(n))$ of the process such that the conditional
expectation of its increments are less than a function
$g(n,\omega)$ multiplying the learning parameter $\eta(n)$, then
$\omega(n)$ converges to $B$, if the function $g$ is negative in a
certain spherical neighborhood of $B$ and if the learning
parameter decreases not so quickly. So it is enough  that
\[\lim_{n\rightarrow+\infty}\eta(n)=0\] in order that the a.e. convergence of
the weights holds. The interesting fact is that the stronger
condition
\[\sum_{n=1}^{+\infty}\eta(n)^2<+\infty . \]
is not introduced. The result of this theorem is neat because the
condition
\[\lim_{n\rightarrow+\infty}\eta(n)=0\] is the one used in the
numerical applications. In Section 3 we will give many examples of
"good" and "bad" decay of $\eta(n)$. The choice of $\eta(n)$ is
important also for the speed of convergence of the process.
Another key role for the a.e. convergence is the form of the
probability distribution of the data as it will be clear from the theorem we
present below.\\
In order to understand it we need other definitions. Let us
introduce a function $g$ which is the leading term of the super
martingale difference given in the proof of theorem
\ref{teorema1}. It is a particular realization of the function $g$
used in the theorem of Nevel'son and Has'minskii:
\begin{equation}\label{g}
g(y_1,y_2,...,y_N;\omega_1,\omega_2,...\omega_N)=\sum_{i=1}^N(y_i-\omega_i)
\cdot \sum_{k}(\int_{\Pi(y)_k}\Lambda(k,i)(x-y_i)f(x)dx).
\end{equation}
where $\omega_i(n)=(\omega_{ij}(n),j=1,...M, i=1,...,N )$ and
$y_i(n)=( y_{ij}(n),j=1,...M, i=1,...,N)\in\mathbb{R}^{M\times N}$
$f$ is the density of the probability distribution of the data
with support on a compact set $\Omega$ of $\mathbb{R}^M$, $\Pi(y)$
is the Voronoi tessellation associated with $y$ (see
(\ref{tassel})).
$(y_i-\omega_i)\cdot(x-y_i)$ is the M-dimensional scalar product.\\
\\We define also:
\[\Theta\equiv\{the \; set \;of\; all\; Voronoi\; tessellations\; associated\;
with\; \{\omega_1(n),...,\omega_N(n)\} \; \;\;for \,all \; n\}
\] For $y\in\mathbb{R}^M$ we use the convention that  $y\in\Theta$
implies that there exists a Voronoi tessellation $\Pi(y)$ such that $\{\Pi(y)_i,i=1...N\}\in\Theta$.
Finally we can enunciate our theorem:\\

\begin{Theorem}\label{teorema1}

Let the vectors $\omega(n) \in \mathbb{R}^{M \times N}$ be updated
by the Kononen algorithm (\ref{2.2})
 \[  \; \;\; \omega_i(n+1)=\omega_i(n)+\eta(n)\Lambda(i,v)\mathrm{\overline{I}}(\omega_v(n),\xi(n+1))\
  \cdot\ (\xi(n+1)-\omega_i(n))\]
if there exists a unique point $\tilde \omega=(\tilde
\omega_1,...\tilde \omega_N)\in\mathbb{R}^{M \times N}$such that
for each $y=(y_1,y_2,...,y_N)$:
\begin{equation}\label{cond1}
 g(y_1,y_2,...,y_N;\tilde \omega_1,...\tilde \omega_N)\leq 0 \
\forall y\in\Theta
\end{equation}
where the equality holds if and only if $ y_i=\tilde \omega_i \
i=1,...N$ and :
\begin{equation}\label{cond2}
\sum_{n=1}^{+\infty}\eta(n)=+\infty \, \ \ \
\lim_{n\rightarrow+\infty}\eta(n)=0
\end{equation}
 then we almost everywhere have:
\[
\lim_{n\rightarrow+\infty}\omega_i(n)=\tilde \omega_i \ \ \ \
i=1,...,N
\]
\end{Theorem}

\begin{remark}
This theorem is interesting because the rate of decay of $\eta(n)$
is the one used in simulations but it is still not enough because
the full proposition should exclude the decays which are not used
in the simulations i.e. the ones such that
\[\sum_{n=1}^{+\infty}\eta(n)^2<+\infty . \] This last condition
is often required in the proofs of theorem about the convergence
of Kohonen algorithm, but we have checked in our simulation that
there is no convergence. For example if we use
$\eta(n)=\frac{1}{n}$ the limit values of weights are not ordered
at the end of the learning process for any initial condition (that
is for any random choice of weights at the beginning of the
algorithm). This result contradicts that one of Sadeghi
(\cite{[23]}). In his paper he made a numerical check but it is
not enough since he has proven directly only the convergence in
mean and not the a.e. convergence and in addition in his
simulation he started from ordered weights.
\end{remark}

\begin{remark}
Although the theorem is formulated in the multi-dimensional case
we use it in one dimension because the condition (\ref{cond1}) is
not easy to check in the general case. For $M=1$ it has been seen
in the paper (\cite{[25]}) that, if the distribution of the data
is uniform and the data belong to the interval $(0,1)$, the
clusters are intervals of amplitude $0.1$ for $N=10$. They are
centered around the points $(0.5, 1.5,.....)$. If the data are
gaussian distributed, as in the biological case, there is no
unique point satisfying condition (\ref{cond1}) and other
arguments must be used. We show in Section 3 that, choosing
$\eta(n)$ in a particular way, it is still possible to have a.e.
convergence but there is no theorem justifying this result.
\end{remark}
\textsl{\textbf{Proof}}
\\ \\The proof goes like in the paper
\cite{[25]}. Let $B$ be the point $\tilde\omega$ of the theorem,
$U_{\varepsilon}(B)$ be the $\varepsilon $ spherical neighborhood
of $\tilde \omega$, $\tau(\epsilon)$ the first $n$ for which the
process $\omega(n)$ enters in $U_{\varepsilon}(B)$. Let $\sigma_n$
be the stopping time
\[\sigma_n\equiv \tau(\varepsilon)\wedge
n=\min(n,\tau(\varepsilon)).\] The function $W(n, \sigma_n)$ of
the theorem of Nevel'son and Has'minskii for the case of the
Kohonen algorithm is

\[ W(n,\omega(n))=\sum_{i=1}^N\|\omega_i(\sigma_n)-\tilde \omega_i\|^2 \]
In effect the condition (\ref{condoper1}) on the function $W(n,
\omega(\sigma_n))$ is nothing other than the non negative super
martingale condition, so if it is possible to show this condition
it is possible to apply the convergence property of martingales.\\
So we start proving:
\begin{equation}\label{teo1}
\sum_{i=1}^N[\mathbb{E}(\|\omega_i(n+1)-\tilde\omega_i\|^2|\mathcal{F}_n)-\|\omega_i(n)-\tilde\omega_i\|^2]\leq
0
\end{equation}
The details of the proof can be found in (\cite{[25]}) here we give the main results\\
\[LW(n,\omega(n))=\mathbb{E}(\;W(n+1,\omega(n+1))-W(n,\omega(n))|\mathcal{F}_n\;)= \]
\[=\sum_{i=1}^N
\mathbb{E}(\|\omega_i(\sigma_{n+1})-\tilde\omega_i\|^2|\mathcal{F}_n)-\|\omega_i(\sigma_n)-\tilde\omega_i\|^2]=\]
\begin{equation}\label{teo5}
=\sum_{i=1}^N\eta(n)(\omega_i(\sigma_n)-\tilde\omega_i)\cdot\sum_v\int_{\Pi(\omega(\sigma_n))_v}\Lambda(v,i)(x-\omega_i(\sigma_n))f(x)dx
+\eta^2(n)g_1(\omega(\sigma_n))
\end{equation}
 Where:
\[g_1(\omega(\sigma_n)=\sum_{i=1}^N \sum_v \int_{\Pi(\omega(\sigma_n))_v}\Lambda(v,i)\|x-\omega_i(\sigma_n)\|^2f(x)dx\]
with $\omega(n)=(\omega_1(n),\omega_2(n),...,\omega_N(n)))$\\
 But
\begin{eqnarray*}
g_1(\omega(\sigma_n)\leq\\
&\leq&\sum_{i=1}^N \sum_v \int_{\Pi(\omega(\sigma_n))_v}|\Lambda(v,i)\|x-\omega_i(\sigma_n)\|^2f(x)|dx\\
&\leq&\sum_{i=1}^N \sum_v \int_{\Pi(\omega(\sigma_n))_v}A\ f(x) dx
\end{eqnarray*}
\begin{equation}\label{teo6}
  =N\cdot a \cdot A\equiv\widetilde{A}
\end{equation}
since $|\Lambda(v,i)|\leq a $, $a \geq 0$, and where A is a
positive constant such that :
\[ \max\{\|x-\omega_i(\sigma_n)\|^2\}\leq A\]
so for (\ref{teo6}),and the conditions (\ref{cond1}) and
(\ref{cond2}) we obtain:
\begin{equation}\label{teo7}
\lim_{n\rightarrow+\infty}\frac{\eta^2(n)\ g_1(\omega(\sigma_n))
}{\eta(n)\ g(\omega(\sigma_n))}=0
\end{equation}
Hence, for n large enough, the sign of the term (\ref{teo5}) is
determined by the sign of $g(\omega(\sigma_n))$ and so we have:
\[\sum_{i=1}^N[\mathbb{E}(\|\omega_i(\sigma_{n+1})-\tilde \omega_i\|^2|\mathcal{F}_n)-\|\omega_i(\sigma_{n})-\tilde\omega_i\|^2]\leq0\]
From this inequality it follows that
\[
W(n,\omega(\sigma_n))\equiv\sum_{i=1}^N\|\omega_i(\sigma_n)-\omega_i\|^2
\]
is a non negative super-martingale .\\
Since $W(n,\omega(\sigma_n))$ is a non negative super-martingale,
from the theorem about martingale the limit exists almost
everywhere, in addition, by the definition of the \emph{stopping
time} $\sigma_n$ and assuming that
 $\tau(\varepsilon)<\mathrm{C}(\varepsilon)$  we have that \\
\[\exists\ C>0 \; \;
such \;that\;\;\ \lim_{n\rightarrow+\infty}W(n,\omega(\sigma_n))=C
\;\; a.e.\]
Hence we found the main inequality of the theorem of Nevel'son and Has'minskii :\\
\begin{equation}\label{teo10}
  LW(n,\omega(n))\leq \eta(n)g(\omega(n))
\end{equation}
\\
From  (\ref{cond1}) we get that (\ref{condoper2}) holds
\begin{equation}\label{teo11}
  \inf_{\stackrel{n\geq Q}{\omega(n)\in V_{\varepsilon,R}(B)}}\
  [-g(\omega(n))]>0.
\end{equation}
In addition
\begin{equation}\label{teo9}
  \inf_{n\geq0}W(n,\omega(n))\longrightarrow+\infty \;
  \;\;  \mbox{per}|\omega(n)|\rightarrow+\infty
\end{equation}
Thus we can apply the theorem of Nevel'son and Has'minskii, where
$Q=Q(\varepsilon,R),(\ R>\varepsilon>0)$ is some constant,
$\varepsilon \;$ is a small enough parameter, $ \;
B_\varepsilon=\{x \in\mathbb{R}^{N\times M} \mbox{such that}
\|x-\tilde\omega\|<\varepsilon\}$ is a spherical neighborhood of
the limit point, \; $B=\bigcap_{\varepsilon>0} B_\varepsilon$,
$U_\varepsilon(B)=\{\omega :
\rho(\omega,B)<\varepsilon\}$\\
$V_\varepsilon(B)=\mathbb{R}^{N\times M}\backslash
U_\varepsilon(B)$,
$V_{\varepsilon,R}(B)=V_\varepsilon(B)\cap\{(\omega(n)
:|\omega(n)|<R\}$  \\ \\
Considering the above statements (\ref{teo10}, \ref{teo11},
\ref{teo9} and \ref{cond2}) we note that the hypothesis
of theorem \ref{teoremaoper} are satisfied and so we obtain\\
\begin{equation}\label{teo12}
\mathrm{P}\{\liminf_{n\rightarrow+\infty}\ \rho(\omega(n),B)=0\}=1
\end{equation}
where $\rho(\omega(n),B)=\inf_{y\in B}\rho(\omega(n),y)$\\ \\
Now by (\ref{teo12}) we have that when $n\rightarrow+\infty \;\;\;
\omega(n)\rightarrow \tilde \omega$ with
probability 1.\\
In fact, since
\[
\lim_{\omega(n)\rightarrow\omega}\ [\sup
_{n\geq0}W(n,\omega(n))]=0\] we have
\[\mathrm{P}\{\lim_{n\rightarrow+\infty}\ W(n,\omega(n))=0\}=1
\]
Thus we get:
\[
\lim_{n\rightarrow+\infty}\omega_i(n)=\tilde \omega_i \ \ \ \
i=1,...,N \ a.e.
\]
In addition as it has been proven  in (\cite{[25]}), the algorithm
will achieve the given accuracy $\varepsilon$ within a finite
number of updates, that is
$\tau(\varepsilon)<\mathrm{C}(\varepsilon)$.

\section{ Numerical studies }

In this section we illustrate our numerical simulations about the
convergence of the Kohonen algorithm. First we consider a
uniformly distributed data set, then a normal distributed data
set, all the data are one dimensional as we already said.\\We see
that the algorithm does not even converge in mean ( and so also
not a.e.) if:
\begin{enumerate}
    \item $\eta(n)$, the learning parameter, decreases too fast
    \item The neighborhood functions ( $ \Lambda(i,v)$ (\ref{2.3}) , or $ h(i,v,n)$
    (\ref{2.4}) ) have a range of action  too small or too large.
\end{enumerate}
In addition, although the learning parameter and the neighborhood
function are optimally chosen, the convergence of the algorithm is
 slow and it needs a large number of iterations in order to have a good accuracy.
  So, when the data set is not large enough, it is useful to repeat the presentation of
data several times in random order until we have a large data set. \\
In particular in the case of uniformly distributed data, chosen
inside the interval $[0,1]$, we verify numerically that, having a
large data set, choosing any neighborhood function and using as
learning parameter $ \eta(n) = \frac{1}{n^\alpha}$ with
$\alpha\geq 1$ the algorithm does not converge in any sense ( for
different initial choices of weights we have different outputs)
and the weights are not ordered during the learning procedure .
Instead using $\eta(n) =\frac{1}{n^\alpha}$ , with $\alpha \leq
\frac{1}{3}$ we have the convergence in mean. So the convergence
property depends on the velocity of decay of $\eta(n)$. In fact if
$\eta(n)$ decreases too fast, e.g. $\eta(n) =\frac{1}{n^\alpha}$ ,
with $\alpha\geq 1$, the updated weights change their values very
little during the learning and so the algorithm is not able to
find the final configuration of weights. $\eta(n) \sim 1/n$ is a
too fast decay because after $100$ iterations already the
variation of the weights is very small and so there is no
convergence while $\eta(n) \sim \frac{1}{\sqrt[3]{n}}$ decreases
less quickly ( it assumes values less than 0.01 from
$n>10^6 $ ) and its velocity of decrease is sufficient to have the convergence.\\
The choice of $\eta(n)$ is basic not only for the convergence but
also for accuracy. In fact we can have the convergence of the
algorithm though the algorithm is not able to identify all the
limit weights but only some of them. This happens when the weights
are updated too fast in the last part of the learning procedure or
when the range of $\eta(n)$ does not cover all the interval
$(0,1)$, for example when the range of $\eta(n)$ is $(0.5,1)$.\\We
analyzed the following $\eta(n)$ :
\begin{enumerate}
    \item $\eta(n) = \frac{1}{\sqrt{log(n)}}$
    \item $\eta(n)= \frac{1}{log(n)}$
    \item $\eta(n) = \eta_i( \frac{\eta_f}{\eta_i})^ \frac{n}{nmax}$
    \item $\eta(n) = \frac{ 1}{\sqrt[3]{n}}$\\
 \end{enumerate}(where $\eta_i$ and $\eta_f$ are respectively the initial and
final value of the function $\eta$ and $n_{max}$ the maximum
number of iterations). For all these cases we have convergence in
mean, but for each case there is a different accuracy .\\Choosing
\begin{enumerate}
    \item $\eta(n) = \eta_i(1-\frac{n}{nmax})$
    \item $\eta(n) = \frac{\sqrt{6*log( n)}}{\sqrt(n)+1}$
\end{enumerate}
 we have convergence a.e. The values of the constants and the particular forms of the
 functions $\eta(n)$ have been determined for satisfying the constraint $ 0\le \eta(n)\le 1$.\\
Before explaining the reasons of this statement,
 we want to discuss the connection of the
convergence with the values of the parameters. The choices of the
parameters depend on the data distribution. For example for the
case $3)$, in the case of uniformly and normally distributed data,
generally we have convergence if we choose $\eta_i$ between $0.1$
and $0.9$ and $\eta_f$ between $10^{-6}$ and $0.1$; in the case
$1)$ of the second list the range of $\eta _i $ is $ (0.1, 0.9)$.
Instead for example with log-normal distributed data the range of
$\eta_i$ is $(0.4, 0.9)$ and
 of $\eta_f$ is $(0.01, 0.1)$ in the case $3)$ and in
the other case the range of $\eta _i $ is $(0.4, 0.8)$ .\\
After many simulations we saw that there is convergence in mean
for $\eta(n)$ such that:

 \[\frac{ 1}{\sqrt[3]{n}} \le  \eta(n) \le \frac{\sqrt{6*log( n)}}{\sqrt(n)+1}.\]
instead there is a.e. convergence for $\eta(n)$ such that

\[\frac{\sqrt{6*log( n)}}{\sqrt(n)+1} \le \eta(n) \le \eta_i(1-\frac{n}{nmax})\]
\begin{figure}[!ht]
\begin{center}
\includegraphics[height=5cm,width=12cm]{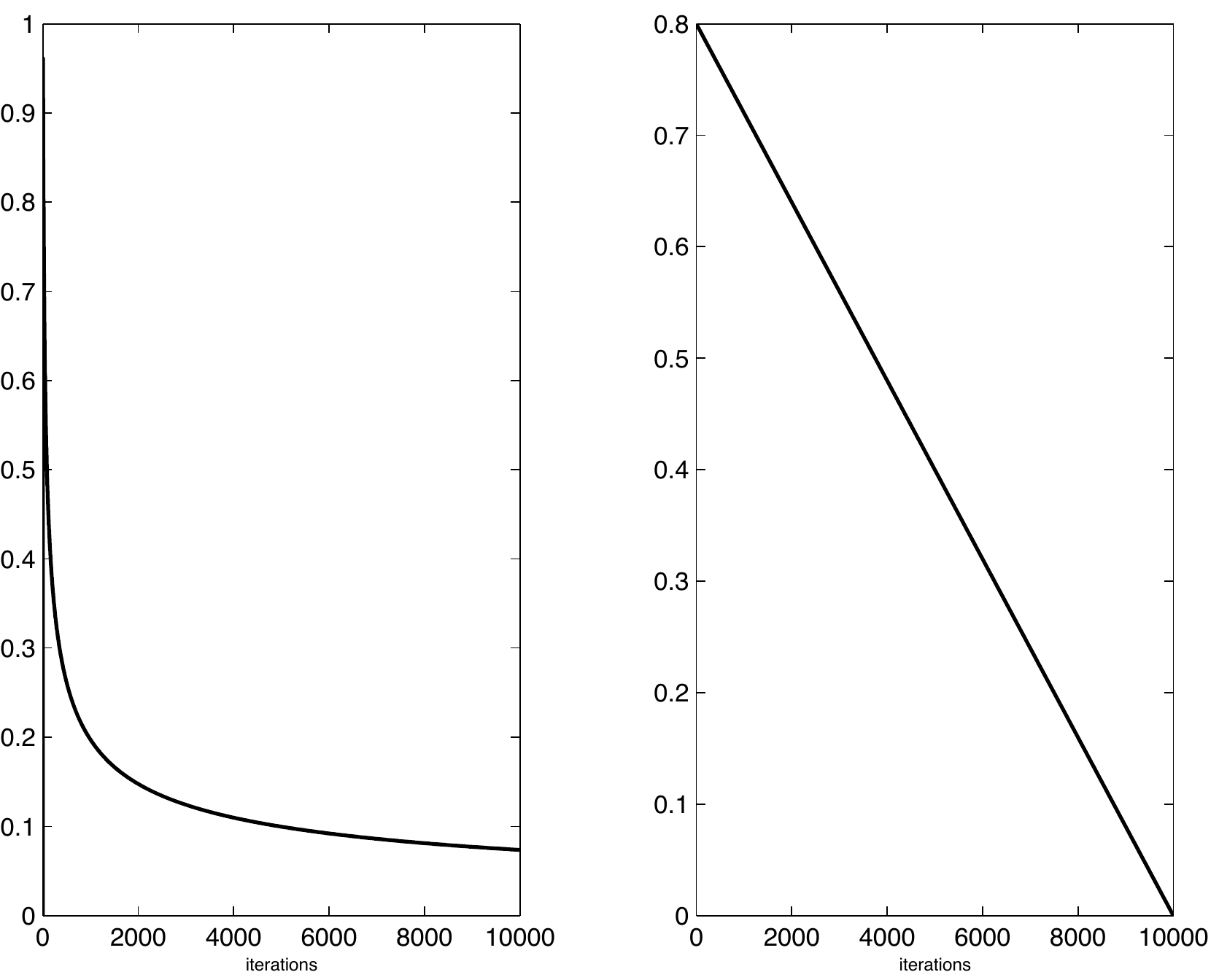}
 \caption{On the left we have  $\eta(n)=\frac{\sqrt{6*log(n)}}{\sqrt(n)+1}$,
 on the right  $ \eta(n) = \eta_i(1-\frac{n}{nmax})$} \label{trend}
\end{center}
\end{figure}The convergence depends also on the values of parameters
concerning the neighborhood function, that is the range of the
action of the winner neuron  which is determined by $s$ in the
case of $\Lambda(i,v)$, and by $\sigma_i$ and $\sigma_f$ in the
case of $ h(i,v,n)$. The choice of $s$ depends strongly on the
number of weights we fix at the beginning and the number of
iterations. For example using a data set of about 10000 uniformly
distributed data if we choose
$\eta(n)=\frac{\sqrt{6*log(n)}}{\sqrt(n)+1}$, $\Lambda(i,v)$ with
$s=1$ as neighborhood function and we want to find 30 groups we do
not have the convergence (the weights are not ordered) but
changing the value of $s$ conveniently (in this case $s \geq 2$) we obtain the convergence.\\
If the data set is smaller than 10000, $s$ is larger than the one
of the previous example.\\ In the case of the h neighborhood
function the best choices of $\sigma_i$ and $\sigma_f$ are the
following:

\begin{equation}\label{sigma1}
   \sigma_i= \frac{\sqrt{N}}{2}
 \end{equation}
 \begin{equation}\label{sigma2}
   \sigma_f= 0.01
 \end{equation}
where $N$ is the number of weights.\\
We have more than one choice for the parameters to obtain the
convergence but different choices give different outputs. We
illustrate some examples. Finding out 10 weights for a data set of
10000 uniformly distributed data, using
$\eta(n)=\frac{\sqrt{6*log(n)}}{\sqrt(n)+1}$ and using h if we
choose $\sigma _i=20$ and $\sigma_f=0.01$ the algorithm converges
and the range of values of weights is $(0.48,0.50)$, in this case
the network identifies $10$ different values inside that interval;
instead if we choose $\sigma _i=5$ and $\sigma_f=0.01$ the range
is $(0.37,0.65)$. We see that the best solution is given by
\ref{sigma1}, \ref{sigma2}, because we have the biggest range of
the weights values, in this case is $0.19$ to $0.82$. It is
important to have the range that covers all the interval of the
data set because otherwise we do not find the optimal partition.
Since we know that for any data distribution the expectation of
weights converges, a small range indicates that the network is
able to find only some of the limit values of weights, in fact for
the reported example,with $N=10$, we know from \cite{[25]} that
the limit values are: $
0.05,0.15,0.25,0.35,0.45,0.55,0.65,0.75,0.85,0.95$, so the range
of the weights values must be $(0.05,0.95)$, more or less. In the
worst choice the network identifies only one limit value. It
happens because the range of action is too big; in this case the
algorithm updates simultaneously too many weights and
they converge to the same value.\\
An analogous situation happens using $\Lambda(i,v)$ as a
neighborhood function. Using $\eta(n)
=\frac{\sqrt{6*log(n)}}{\sqrt(n)+1}$, searching always 10 weights
for a data set of 10000 uniformly distributed data, if we choose $
s\geq 1$ the algorithm converges but the range of weights values
change for different choices of $s$. Increasing $s$ the range of
weights becomes smaller and the weights converge to the same limit
if  $s=10 $. To be more precise if s is equal to $N$, the network
generates $N$ weights ( in this case $N=10$)  with the same value.
The biggest range, in this case, is obtained with $s=1$. If the
number of weights increases the best choice of s is always the
minimum values of s by which we obtain the convergence of the
algorithm. For example in the case
we search 50 weights the best choice is $s=3$.\\
 Summarizing to obtain the convergence we must choose $\eta(n)$
 with a  convenient monotone decay and with a large range; in addition
 we must estimate the right parameters of the neighborhood function
such that we have convergence and the maximum range for the
weights values  in order to determine the optimal partition of data set.\\
As we said previously the error of the expectation of the weights
varies for different choices of $\eta(n)$, and for some choices of
$\eta(n)$ we have a.e convergence.\\This statement is based on the
 following analysis: we run the Kohonen algorithm $1000$ times
  for different data sequences. We use at the beginning a set of uniformly
 distributed data of $4000$
  elements , then $10000,20000,30000,60000,120000,150000$ and $250000$.
  This procedure has been done with all the mentioned $\eta(n)$ and both
$\Lambda$ and h.\\At the end of algorithm running for each data
set we have 1000 cases of weights limit values. The mean value of
these cases actually converges to the centers of the optimal
partition of the interval $(0,1)$ for all $\eta(n)$ and for each
neighborhood function.\\ In addition the average error of limit
weights, with respect to the exact values of the centers,
decreases on increasing the number of iterations for
$\eta(n)=\frac{\sqrt{6*log(n)}}{\sqrt(n)+1}$
 and $\eta_i(1-\frac{n}{nmax})$ and any neighborhood
 function; but using  $\Lambda(i,v)$ the error decreases more quickly.\\
Moreover the computing time of the algorithm using h is about 7
times longer than the one  using $\Lambda$ and the accuracy of
weights on the boundaries is worse using h.\\
The weights near the border are not updated symmetrically and so
they are shifted inward by an amount of the order of
$\frac{1}{2*N}$, where $N$ is the number of weights in the case of
$\Lambda(i,v)$ while, using h, the weights which
are shifted are 4, two for each boundary.\\
Now we illustrate the quoted results. The following tables
(\ref{tabella1},\ref{tabella2},\ref{tabella3},\ref{tabella4}) show
the evolution of the error. In the first table there are the
average errors of each weight using $\Lambda$ as neighborhood
function and 10000 uniformly distributed inputs; instead in the
third table there are 60000 uniformly distributed inputs. In the
second and in the fourth table it is shown the case of $h$ as
neighborhood function. The weights are $N=10$, so the limit values
, which are written in the firs column of every tables, are:
0.05,0.15,0.25,0.35,0.45,0.55,0.65,0.75,0.85,0.95
\begin{table}[!h]
\begin{center}\small
\begin{tabular}{|c|c|c|c|c|c|c|c|c|c|c|}\hline
 \\
 &$\eta_i(\frac{\eta_f}{\eta_i})^ \frac{n}{nmax}$&$\frac{1}{\log(n)}$&$ \frac{1}{\sqrt{log(n)}}$&$\eta_i(1-\frac{n}{nmax})$& $\frac{\sqrt{6*log(
 n)}}{\sqrt{n}+1}$\\
 \hline\\
0.05&0.0561& 0.0562 & 0.0576&  0.0559&  0.0559\\
0.15&0.0189 & 0.0197&  0.0358&  0.0088&  0.0160\\
0.25&0.0239 & 0.0246 & 0.0396 & 0.0183 & 0.0216\\
 0.35& 0.0185&  0.0191 &0.0364 & 0.0104 & 0.0158\\
0.45&0.0202 & 0.0211 & 0.0385 & 0.0104 & 0.0168\\
0.55& 0.0189 & 0.0196 & 0.0366 & 0.0107 & 0.0161\\
0.65& 0.0198 & 0.0208 & 0.0389  &0.0111&  0.0167 \\
0.75&0.0238 & 0.0243 & 0.0382&  0.0185&  0.0216 \\
0.85&0.0180 & 0.0188 & 0.0351 & 0.0089 & 0.0158 \\
0.95&0.0555 & 0.0556 & 0.0568 & 0.0558 & 0.0557\\
 \hline
\end{tabular} \caption{Mean error of each weight for 10000
iterations using $\Lambda$ as neighborhood function.The $\eta$ are
shown in the first row. }\label{tabella1}
\end{center}
\end{table}

\begin{table}[!h]
\begin{center}\small
\begin{tabular}{|c|c|c|c|c|c|c|c|c|c|c|} \hline
\\
  &$\eta_i(\frac{\eta_f}{\eta_i})^ \frac{n}{nmax}$&$\frac{1}{\log(n)}$&$ \frac{1}{\sqrt{log(n)}}$&$\eta_i(1-\frac{n}{nmax})$& $\frac{\sqrt{6*log( n)}}{\sqrt{n}+1}$\\

\hline\\
0.05&0.1063&  0.1061&  0.1045&  0.1059&  0.1061\\
0.15&0.0565 & 0.0564 & 0.0587 & 0.0557 & 0.0562 \\
0.25&0.0339  &0.0342 &0.0440& 0.0305 & 0.0325\\
0.35& 0.0262 & 0.0267&  0.0409&  0.0197 & 0.0238\\
0.45& 0.0214& 0.0222 & 0.0388 & 0.0109&  0.0183\\
0.55&  0.0213& 0.0223& 0.0379 &0.0106& 0.0185\\
0.65&   0.0256&  0.0264&  0.0403 & 0.0189& 0.0232\\
0.75&0.0336& 0.0343& 0.0460 & 0.0302 & 0.0322\\
0.85& 0.0567&  0.0570 & 0.0615& 0.0560& 0.0566\\
0.95&  0.1067&  0.1069 & 0.1065 & 0.1064 & 0.1068
 \\
 \hline
\end{tabular}
\caption{Mean error of each weight for 10000 iterations using $h$
as neighborhood function.The $\eta$ are shown in the first
row.}\label{tabella2}
\end{center}
\end{table}

\begin{table}[!h]
\begin{center}\small
\begin{tabular}{|c|c|c|c|c|c|c|c|c|c|c|}\hline
 \\
 &$\eta_i(\frac{\eta_f}{\eta_i})^ \frac{n}{nmax}$&$\frac{1}{\log(n)}$&$ \frac{1}{\sqrt{log(n)}}$&$\eta_i(1-\frac{n}{nmax})$& $\frac{\sqrt{6*log(
 n)}}{\sqrt{n}+1}$ \\
 \hline\\
0.05&0.0565&  0.0563&  0.0578&  0.0560&  0.0563 \\
0.15&0.0190 & 0.0179& 0.0342& 0.0071 & 0.0114\\
0.25& 0.0239 & 0.0234 & 0.0354& 0.0181&  0.0189 \\
0.35 &0.0199& 0.0192&  0.0348 &0.0079 & 0.0117\\
0.45& 0.0206& 0.0194 &0.0365 &0.0073& 0.0113 \\
0.55& 0.0200&  0.0189&0.0384 & 0.0071& 0.0114 \\
0.65&0.0188& 0.0178& 0.0350  &0.0079 & 0.0115\\
 0.75&0.0218& 0.0211&  0.0343&0.0179& 0.0183\\
0.85& 0.0181&  0.0170 & 0.0336 & 0.0070& 0.0108\\
 0.95& 0.0556 & 0.0557& 0.0553& 0.0560& 0.0556\\
 \hline
\end{tabular} \caption{Mean error of each weight for 60000
iterations using $\Lambda$ as neighborhood function.The $\eta$ are
shown in the first row.  }\label{tabella3}
\end{center}
\end{table}

\begin{figure}[!h]
\begin{center}
\includegraphics[height=8cm,width=12cm]{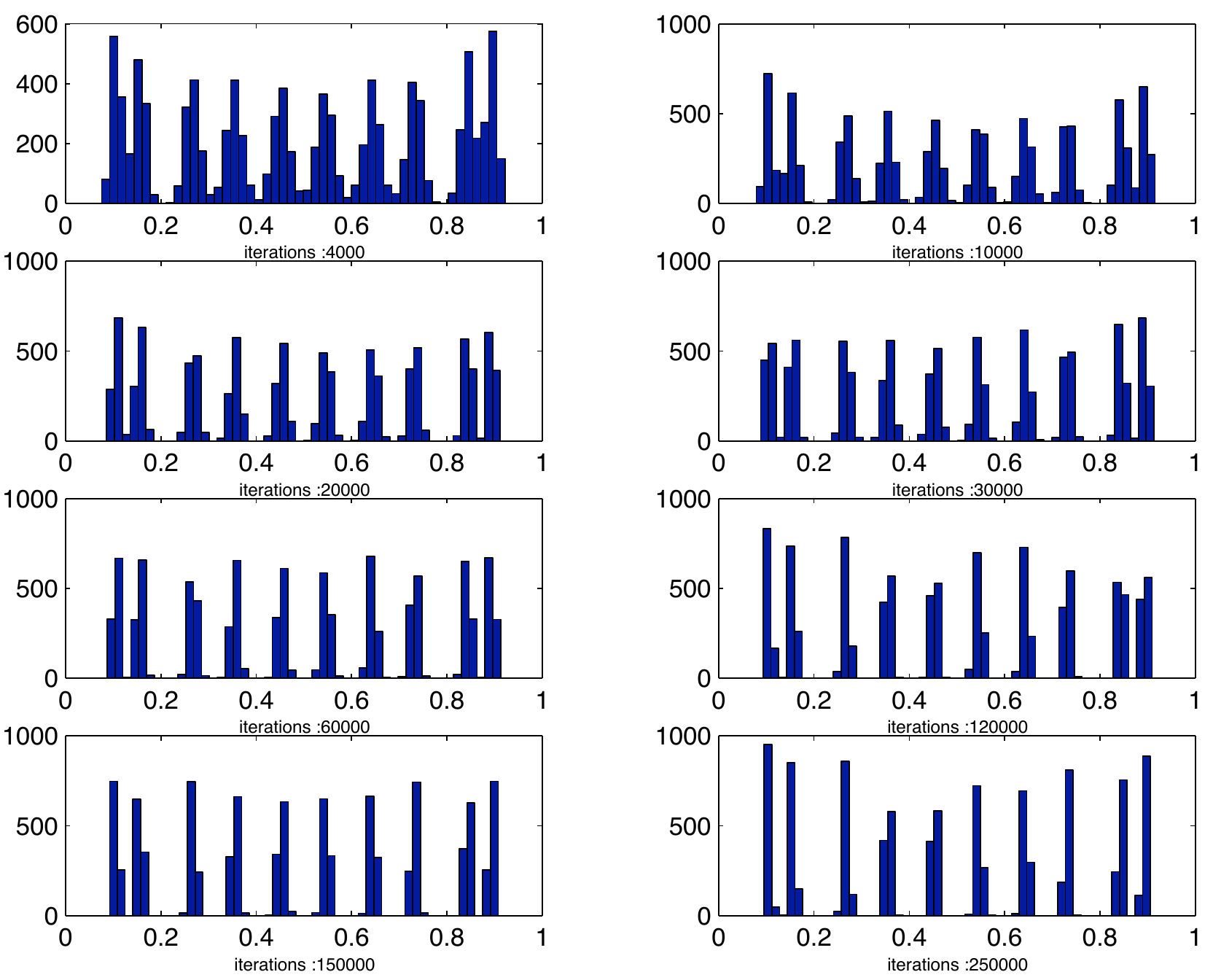}
 \caption{ Histograms in the case of $\eta(n) = \eta_i(1-\frac{n}{nmax})$,
  uniformly distributed data and for different numbers $M$ of iterations } \label{istonmax}
\end{center}
\end{figure}

\begin{figure}[!h]
\begin{center}
\includegraphics[height=8cm,width=12cm]{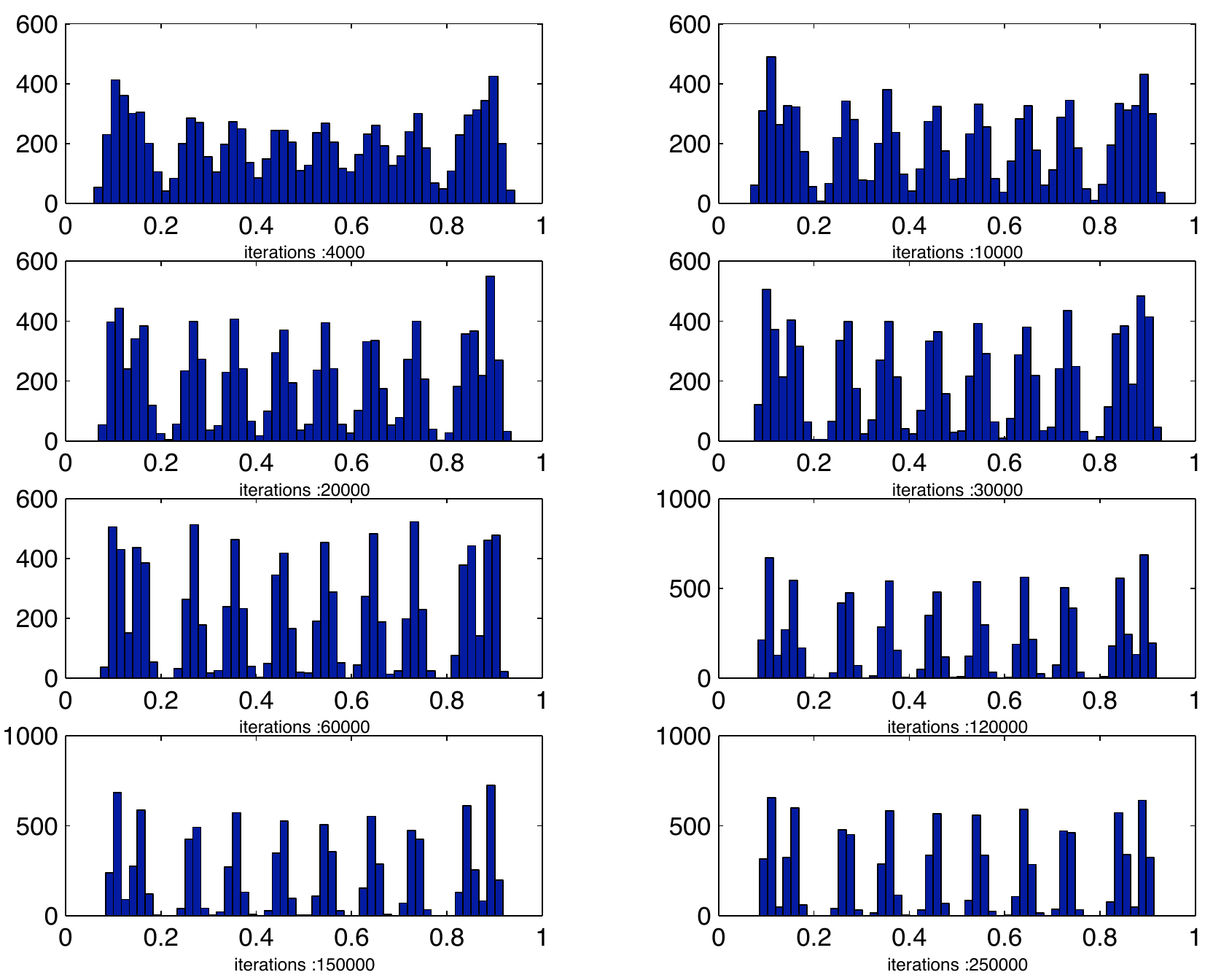}
 \caption{Histograms in the case of $\eta(n) =\frac{\sqrt{6*log(
 n)}}{\sqrt{n}+1}$, uniformly distributed data and for different numbers $M$ of iterations } \label{istolog_rad}
\end{center}
\end{figure}

\begin{figure}[!h]
\begin{center}
\includegraphics[height=8cm,width=12cm]{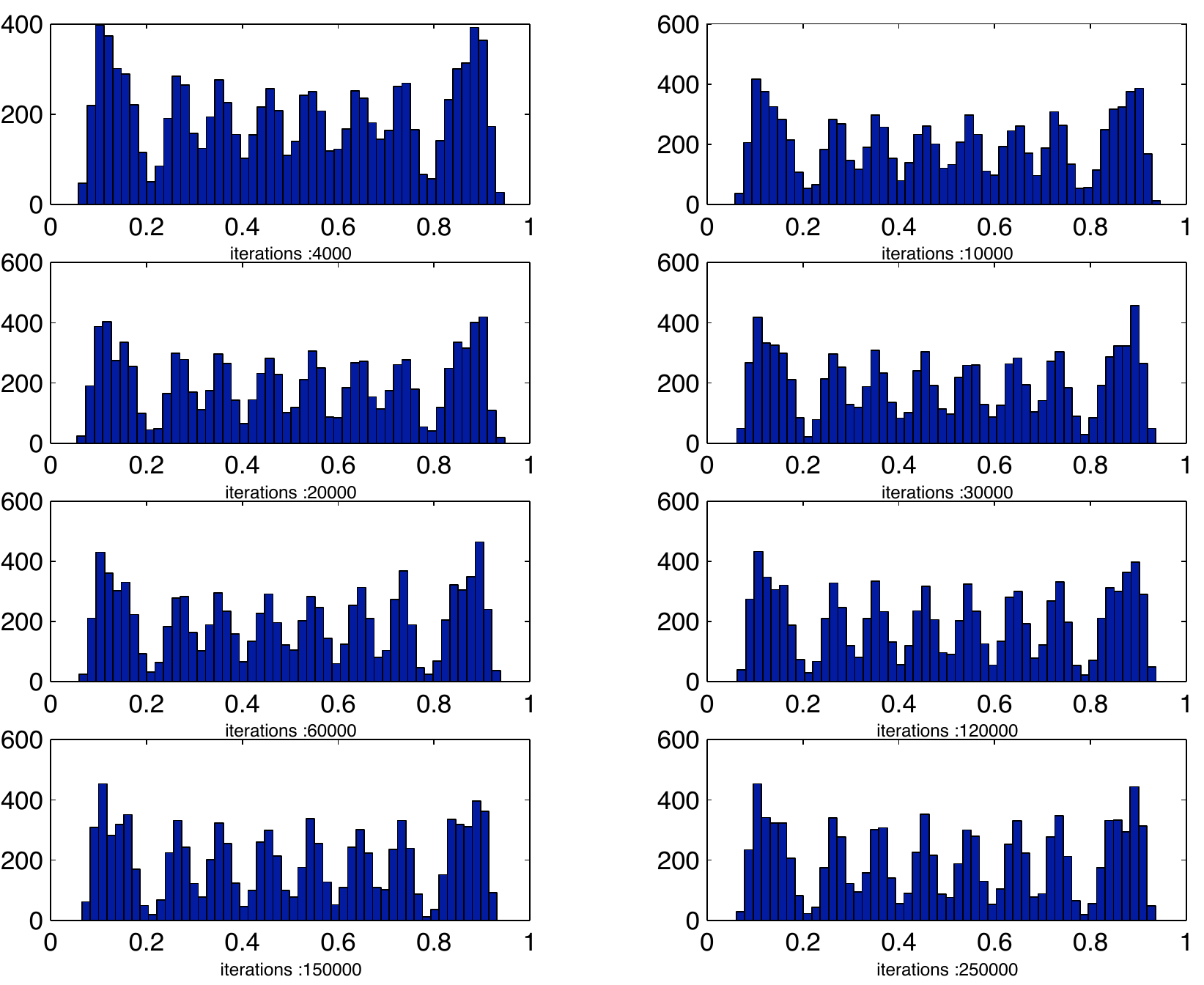}
 \caption{Histograms in the case of $\eta(n) =\frac{1}{log(n)}$,
 uniformly distributed data and for different numbers $M$ of iterations} \label{istolog}
\end{center}
\end{figure}

\begin{figure}[!h]
\begin{center}
\includegraphics[height=8cm,width=12cm]{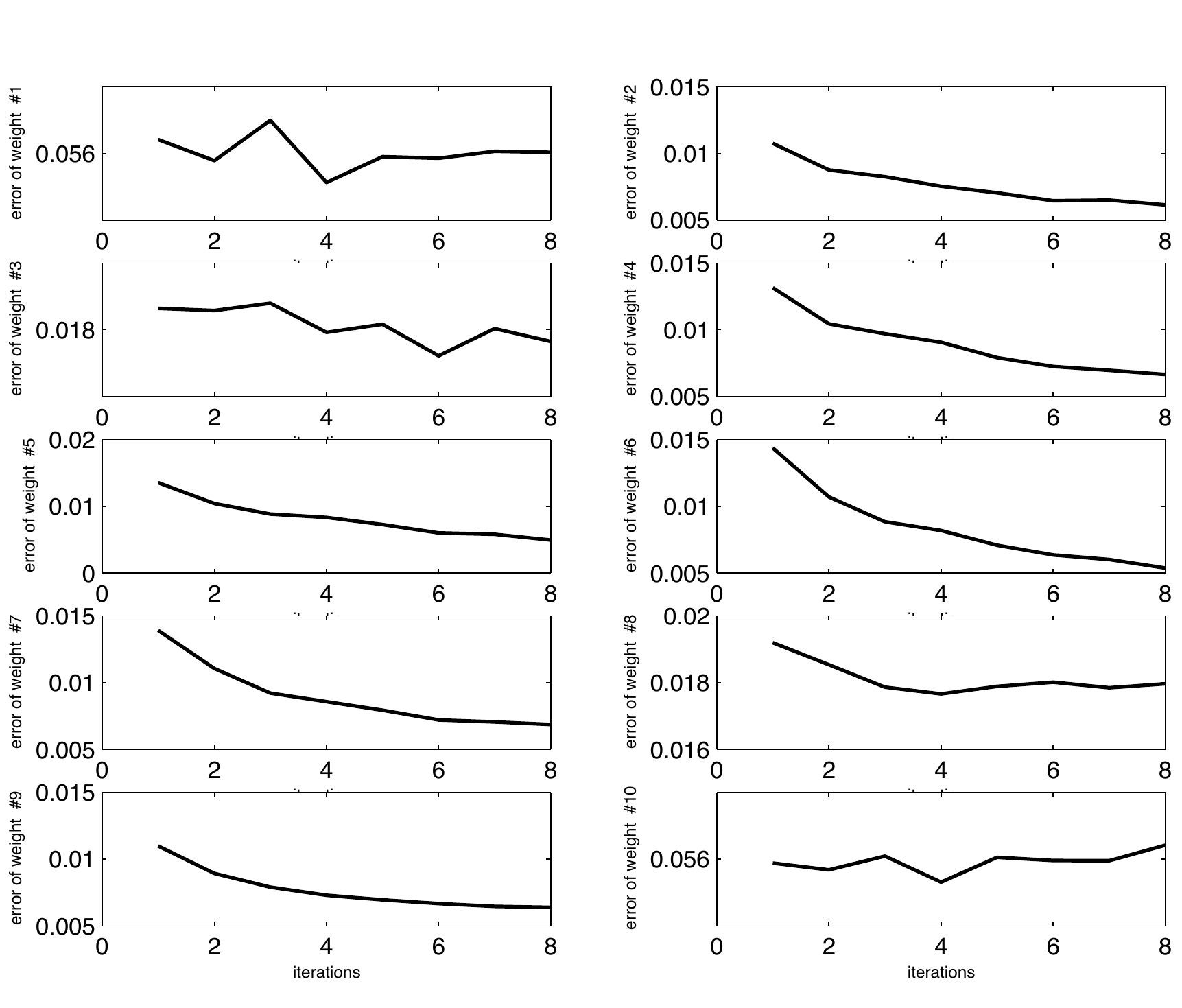}
 \caption{Plot of the mean error for each limit weight in the case of $\eta(n) = \eta_i(1-\frac{n}{nmax})$ and uniformly distributed data.
 The numbers on the x axes indicate the following iterations: $4000$, $10000$, $20000$, $30000$, $60000$, $120000$, $150000$,
 $250000$. } \label{errore_nmax}
\end{center}
\end{figure}

\begin{figure}[!h]
\begin{center}
\includegraphics[height=8cm,width=12cm]{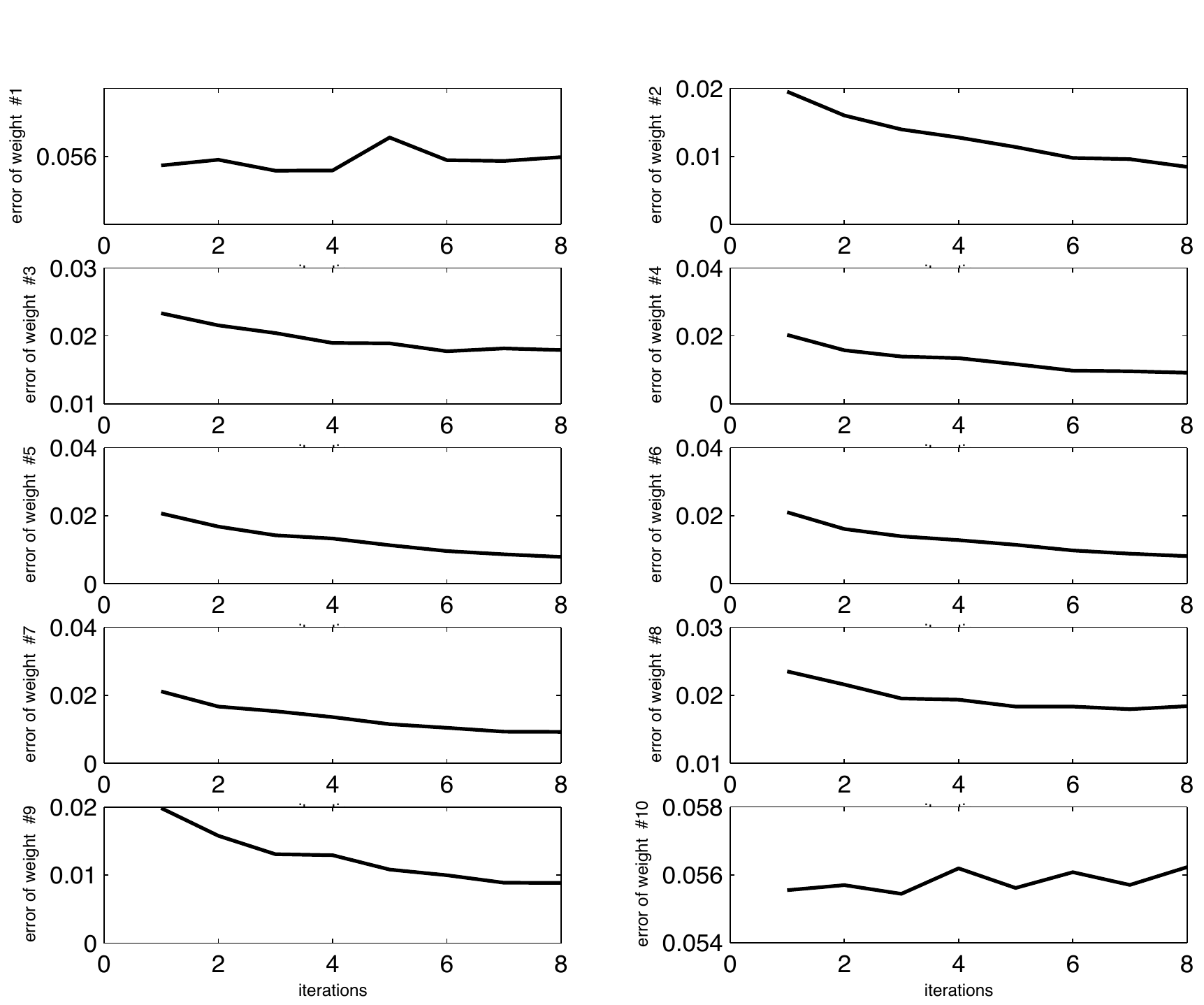}
 \caption{Plot of the medium error of each limit weight in the case of $\eta(n) =\frac{\sqrt{6*log( n)}}{\sqrt{n}+1}$ and uniformly distributed data.
 The numbers on the x axes indicate the following iterations: $4000$, $10000$, $20000$, $30000$, $60000$, $120000$, $150000$,
 $250000$. } \label{errore_log_rad}
\end{center}
\end{figure}

\begin{figure}[!h]
\begin{center}
\includegraphics[height=8cm,width=12cm]{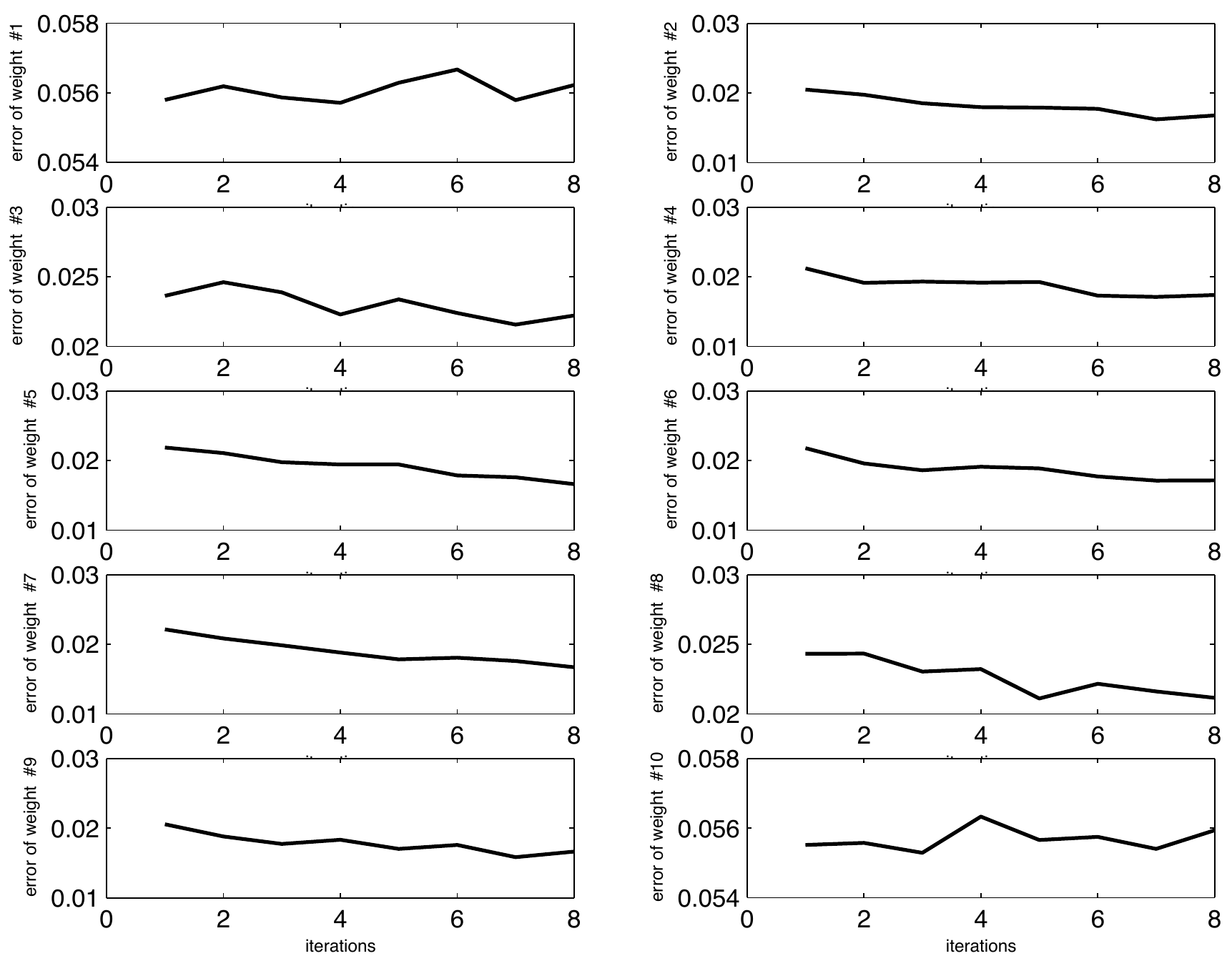}
 \caption{Plot of the medium error of each limit weight in the case of $\eta(n) =\frac{1}{log(n)}$ and uniformly distributed data.
 The numbers on the x axes indicate the following iterations: $4000$, $10000$, $20000$, $30000$, $60000$, $120000$, $150000$,
 $250000$.} \label{errore_log}
\end{center}
\end{figure}

\begin{table}[ht]
\begin{center}\small
\begin{tabular}{|c|c|c|c|c|c|c|c|c|c|c|}
 \hline\\
  &$\eta_i(\frac{\eta_f}{\eta_i})^ \frac{n}{nmax}$&$\frac{1}{\log(n)}$&$ \frac{1}{\sqrt{log(n)}}$&$\eta_i(1-\frac{n}{nmax})$& $\frac{\sqrt{6*log( n)}}{\sqrt{n}+1}$\\
\\ \hline\\
0.05&0.1061 & 0.1060&  0.1040&  0.1064&  0.1059\\
0.15& 0.0559&  0.0558& 0.0580& 0.0560&  0.0555\\
0.25& 0.0326 & 0.0321& 0.0362& 0.0302&  0.0300\\
0.35& 0.0243& 0.0235& 0.0340 &0.0185 & 0.0193\\
0.45& 0.0206& 0.0196 &0.0361& 0.0082& 0.0125 \\
0.55&0.0215 &0.0205& 0.0373 &0.0088& 0.0132\\
0.65& 0.0252& 0.0244 &0.0356& 0.0191& 0.0198\\
 0.75&0.0321 &0.0317&0.0345& 0.0306& 0.0299\\
 0.85& 0.0548& 0.0548& 0.0565& 0.0562& 0.0555\\
 0.95& 0.1051&0.1053& 0.1045& 0.1064 &0.1061\\
 \hline
\end{tabular}
\caption{Mean error of each weight for 60000 iterations using $h$
as neighborhood function.The $\eta$ are shown in the first
row.}\label{tabella4}
\end{center}
\end{table}
We give some examples to illustrate the error evolution using
$\eta(n) = \eta_i(1-\frac{n}{nmax})$ and
$\eta(n)=\frac{\sqrt{6*log( n)}}{\sqrt{n}+1}$, the case of
a.e. convergence.\\
The tables \ref{tabella5} and \ref{tabella6} concern the
application of the algorithm with $4000$, $10000$, $20000$,
$30000$, $60000$, $120000$, $150000$, $250000$ iterations, which
are written in the first column, and using $\Lambda$ as
neighborhood function
\begin{table}[!h]
\begin{center}\small
\begin{tabular}{|c|c|c|c|c|c|c|c|c|c|c|}
\hline\\
$\eta(n)=\eta_i(1-\frac{n}{nmax})$&err1&err2&err3&err4&err5&err6&err7&err8&err9&err10\\
\hline\\
 4000&0.0562&0.0108& 0.0183& 0.0132& 0.0135& 0.0144& 0.0139& 0.0192& 0.0110& 0.0559\\
10000&0.0559& 0.0088 &0.0183& 0.0104 &0.0104& 0.0107&0.0111&0.0185&0.0089& 0.0558\\
20000&0.0565& 0.0083& 0.0184& 0.0097& 0.0089& 0.0088&0.0092&0.0179&0.0079& 0.0560\\
30000&0.0556& 0.0075& 0.0180& 0.0091& 0.0083& 0.0082&0.0086&0.0177&0.0073& 0.0557\\
60000&0.0560& 0.0071& 0.0181& 0.0079& 0.0073& 0.0071& 0.0079& 0.0179&0.0070 &0.0560 \\
120000&0.0559& 0.0065& 0.0176& 0.0072& 0.0060& 0.0064& 0.0072& 0.0180&0.0067 &0.0560\\
150000&0.0560& 0.0065& 0.0180& 0.0070& 0.0058& 0.0060& 0.0071& 0.0178&0.0065& 0.0560\\
250000&0.0560& 0.0062& 0.0178& 0.0067& 0.0050& 0.0054&0.0069&0.0180& 0.0064& 0.0562\\
 \hline
\end{tabular}
\caption{Evolution of the mean error for each weight in the case
of $\eta(n) = \eta_i(1-\frac{n}{nmax})$.$err_i$ means: mean error
of weight i}\label{tabella5}
\end{center}
\end{table}
\begin{table}[!h]
\begin{center}\small
\begin{tabular}{|c|c|c|c|c|c|c|c|c|c|c|}
 \hline\\
$\eta(n)=\frac{\sqrt{6*log( n)}}{\sqrt{n}+1}$&err1&err2&err3&err4&err5&err6&err7&err8&err9&err10\\
\hline\\
 4000&0.0559&0.0195& 0.0233& 0.0203 &0.0207& 0.0210 &0.0211& 0.0235& 0.0198& 0.0556\\
10000&0.0559 &0.0160& 0.0216 &0.0158& 0.0168&0.0161&0.0167&0.0216&0.0158& 0.0557\\
20000&0.0558& 0.0139& 0.0204 &0.0140 &0.0142& 0.0140& 0.0153& 0.0195&0.0131 &0.0554\\
30000&0.0558& 0.0128& 0.0190& 0.0135& 0.0133&0.0128&0.0136&0.0193&0.0129& 0.0562\\
60000&0.0563& 0.0114& 0.0189& 0.0117& 0.0113& 0.0114& 0.0115& 0.0183&0.0108& 0.0556 \\
120000&0.0559& 0.0098 &0.0177& 0.0098& 0.0096& 0.0098& 0.0104& 0.0184&0.0100& 0.0561\\
150000&0.0559& 0.0096& 0.0182& 0.0096& 0.0086& 0.0088& 0.0093&0.0179&0.0089& 0.0557\\
250000&0.0560 &0.0084& 0.0179 &0.0092 &0.0078& 0.0081&0.0092&0.0184 &0.0089& 0.0562\\
 \hline
\end{tabular}
\caption{Evolution of the mean error for each weight in the case
of $\eta(n)=\frac{\sqrt{6*log( n)}}{\sqrt{n}+1}$.$err_i$ means:
mean error of weight i}\label{tabella6}
\end{center}
\end{table}
As seen in the tables the error decreases faster using
$\eta(n)=\eta_i(1-\frac{n}{nmax})$ and it decreases increasing the
iterations; see the figures
\ref{istonmax},\ref{istolog_rad},\ref{istolog} and figures
\ref{errore_nmax}, \ref{errore_log_rad}, \ref{errore_log}.
 In some of these pictures there are the histograms of the limit weights
values obtained running the algorithm 1000 times for different
numbers $M$ of iterations using every time a specific $\eta(n)$.
The histograms show how, increasing $M$, only in the case of
$\eta(n) = \eta_i(1-\frac{n}{nmax})$ and
$\eta(n)=\frac{\sqrt{6*log( n)}}{\sqrt{n}+1}$ the variance of the
histograms tend to $0$, each around a limit value of the weight,
as we expect since we have a.e. convergence. There are only the
histograms for $\eta(n)=\frac{1}{log(n)}$ as example of the
convergence in mean since the other cases are similar.\\ The
velocity of convergence is very slow after 100000 iterations, it
needs many iterations only to change one weight nearer to its
limit value; so to construct the histograms with ten columns we
need a huge data set. In addition in the pictures with the plot of
medium error of each weight we can see that the error decreases
always increasing the iterations only in the case of those
$\eta(n)$ which assure the a.e. convergence. Focusing the
attention on the pictures of the error we see that from 150000
iterations in the case of $\eta(n)= \eta_i(1-\frac{n}{nmax})$ the
error decreases very slowly and for the third and eight weight
there is a little increase of error $\sim 0.002$, it depends on
the propagation of the error of the weights from the boundaries.

\begin{figure}[!h]
\begin{center}
\includegraphics[height=8cm,width=12cm]{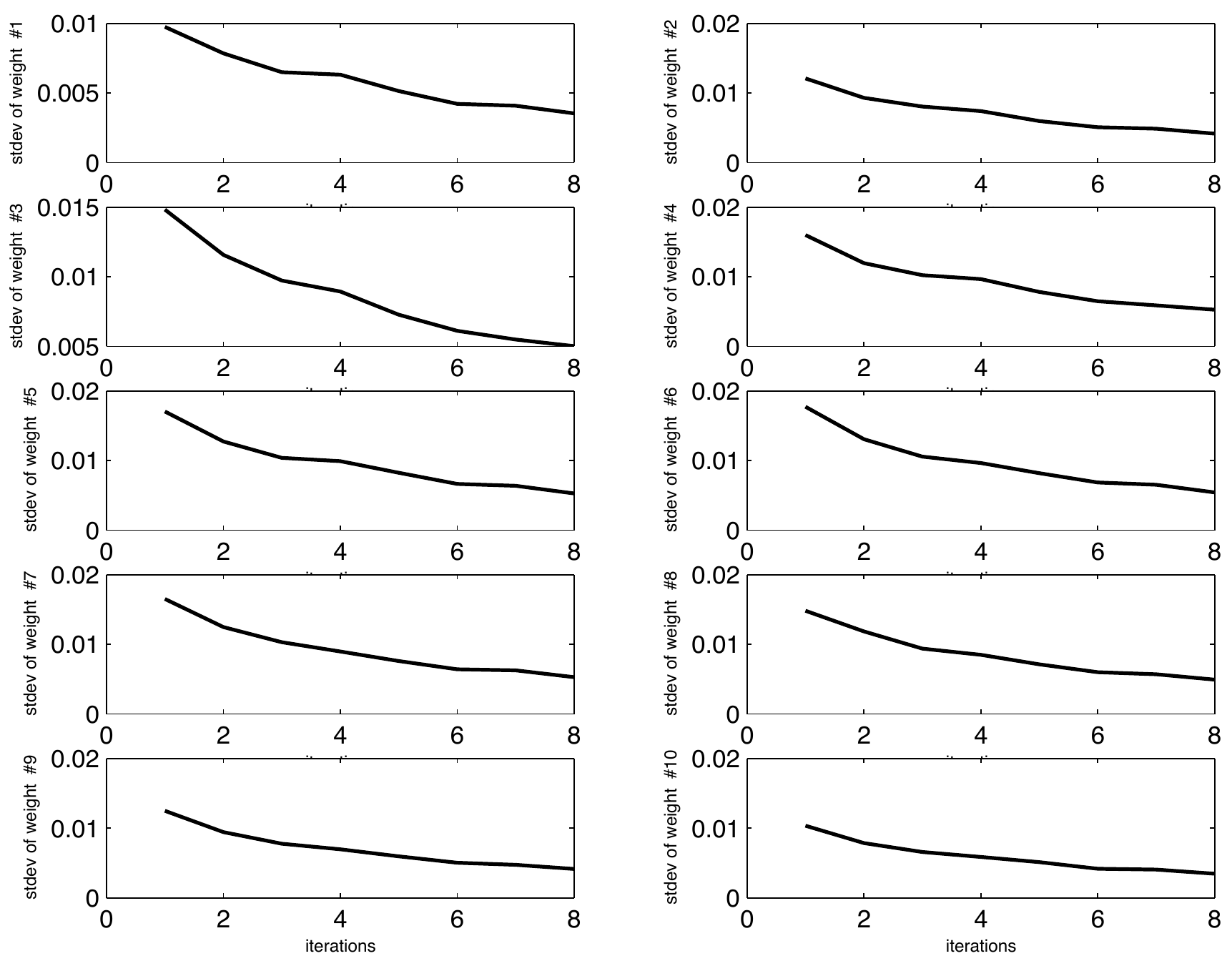}
 \caption{ The standard deviation of the weights in the case of $\eta(n) = \eta_i(1-\frac{n}{nmax})$ and uniformly distributed data.
 The numbers on the x axes indicate the following iterations: $4000$, $10000$, $20000$, $30000$, $60000$, $120000$, $150000$,
 $250000$. } \label{stdnmax}
\end{center}
\end{figure}

\begin{figure}[!h]
\begin{center}
\includegraphics[height=8cm,width=12cm]{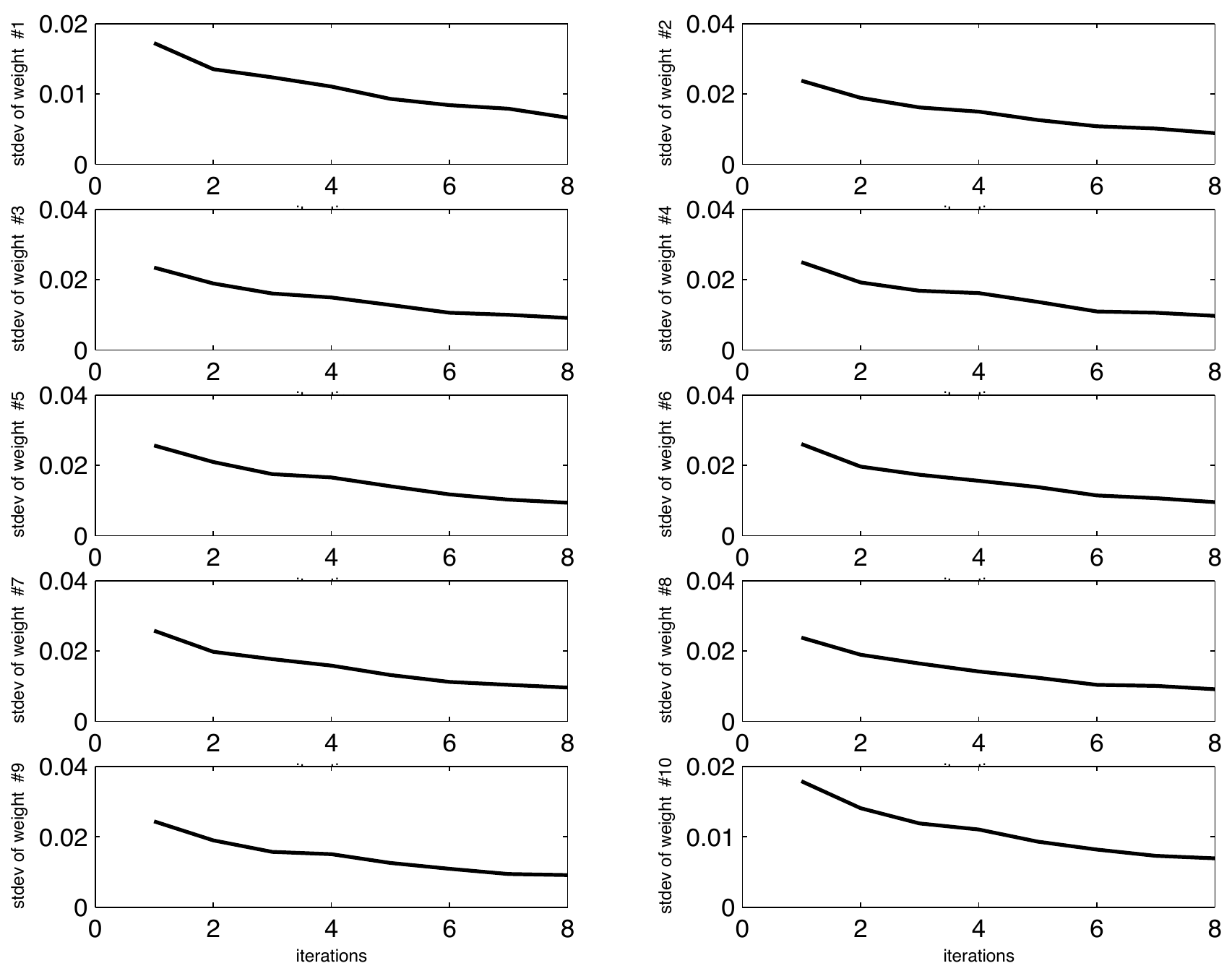}
 \caption{ The standard deviation of the weights in the case of $\eta(n) = \frac{\sqrt{6*log( n)}}{\sqrt{n}+1}$ and uniformly distributed data.
 The numbers on the x axes indicate the following iterations: $4000$, $10000$, $20000$, $30000$, $60000$, $120000$, $150000$,
 $250000$.} \label{stdlog_rad}
\end{center}
\end{figure}

The a.e. convergence is guaranteed in the case of $\eta(n) =
\eta_i(1-\frac{n}{nmax})$ and $\eta(n)=\frac{\sqrt{6*log(
n)}}{\sqrt{n}+1}$  by the monotonically decrease of standard
deviation of weights as the figures \ref{stdnmax} and
\ref{stdlog_rad} show.

\begin{figure}[!h]
\begin{center}
\includegraphics[height=8cm,width=12cm]{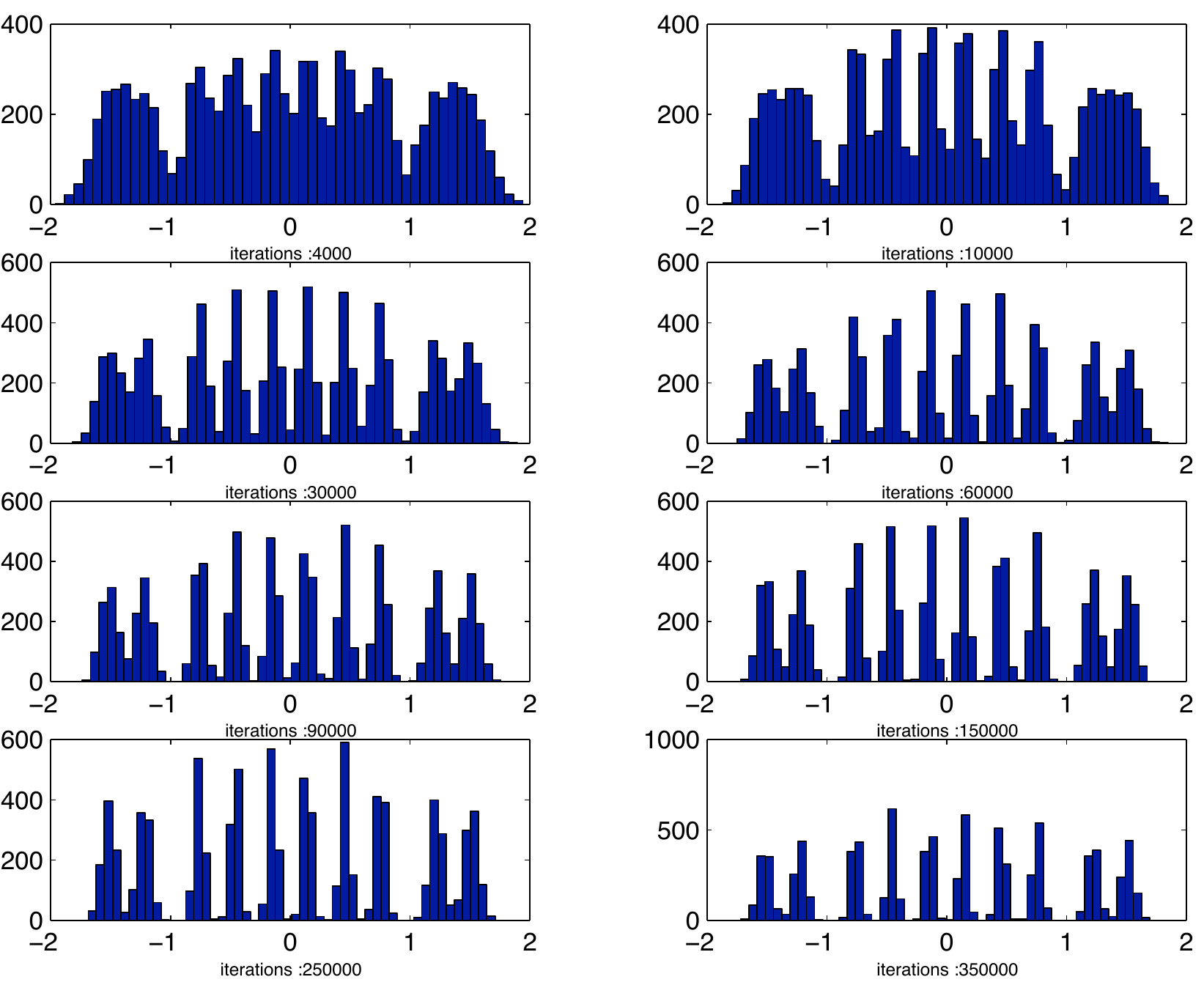}
 \caption{Histograms in the case of $\eta(n) =\frac{\sqrt{6*log( n)}}{\sqrt{n}+1}$,
 normal distributed data and for different numbers $M$ of iterations } \label{istolog_radgauss}
\end{center}
\end{figure}

\begin{figure}[!h]
\begin{center}
\includegraphics[height=8cm,width=12cm]{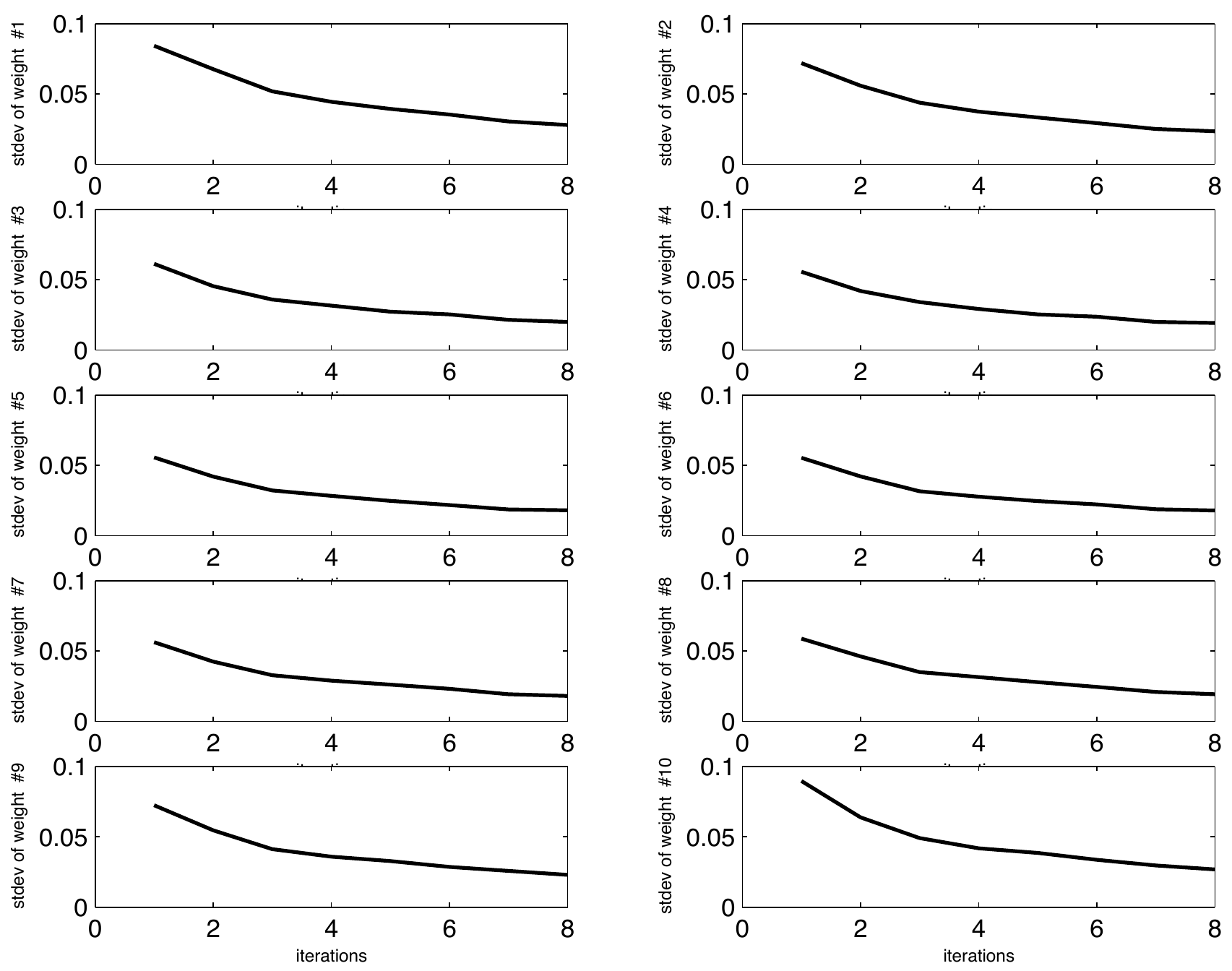}
 \caption{The standard deviation of the weights in the case of $\eta(n) =
 \eta_i(1-\frac{n}{nmax})$ and
  normal distributed data.The numbers on the x axes indicate the following iterations: $4000$, $10000$, $30000$, $60000$, $90000$, $150000$,
 $250000$, $350000$.} \label{stdnmax_gauss}
\end{center}
\end{figure}

\begin{figure}
\begin{center}
\includegraphics[height=8cm,width=12cm]{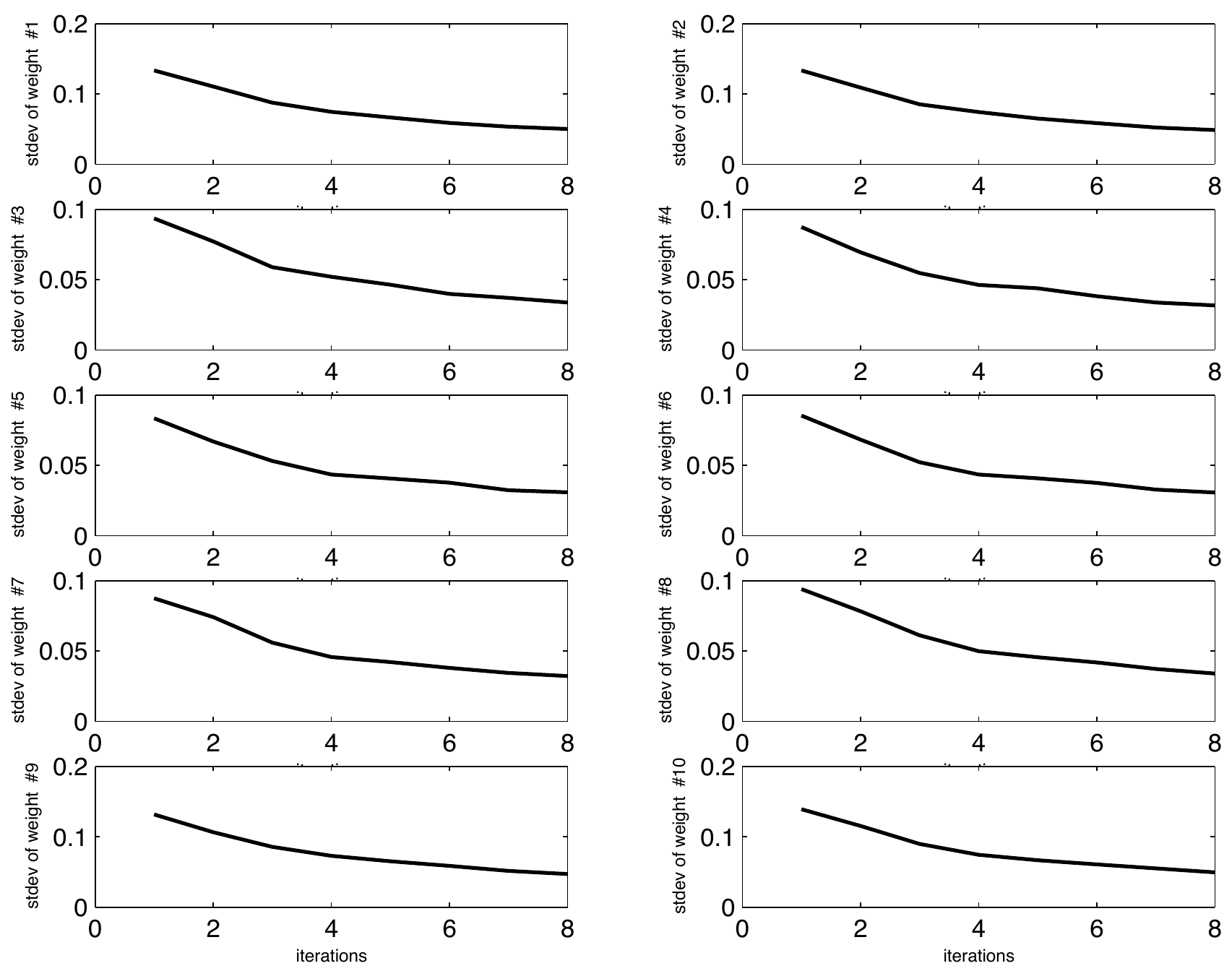}
\caption{The standard deviation of the weights in the case of
$\eta(n)=\frac{\sqrt{6*log( n)}}{\sqrt{n}+1}$ and normal
distributed data.The numbers on the x axes indicate the following
iterations: $4000$, $10000$, $30000$, $60000$, $90000$, $150000$,
 $250000$, $350000$.} \label{stdlog_radgauss}
\end{center}
\end{figure}

Similar results are obtained with the normally distributed data
but for a special choice of the learning parameter. In fact our
theorem does not hold for gaussian
distribution as we already mentioned.\\
Also in this case we have convergence in mean with all the $\eta$
and for any neighborhood function; and convergence a.e for
$\eta(n)=\eta_i(1-\frac{n}{nmax})$ and
$\eta(n)=\frac{\sqrt{6*log(n)}}{\sqrt{n}+1}$ . \\We show in the
following histograms (figures
\ref{istonmax_gauss},\ref{istolog_radgauss}) and plots (figures
\ref{stdnmax_gauss},\ref{stdlog_radgauss}) only the results for
a.e convergence:
\begin{figure}[!h]
\begin{center}
\includegraphics[height=8cm,width=12cm]{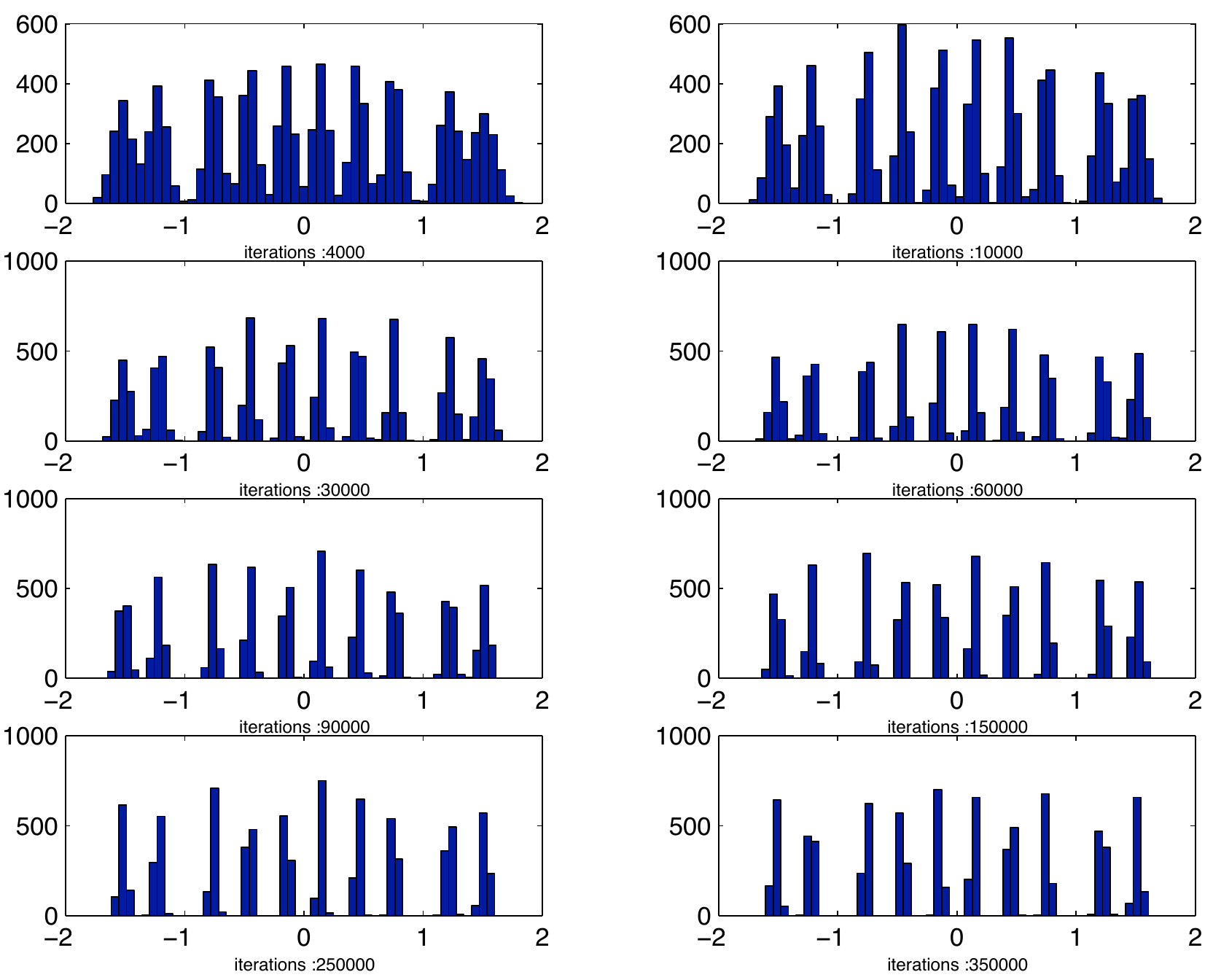}
 \caption{ Histograms in the case of $\eta(n) =
 \eta_i(1-\frac{n}{nmax})$ and
  normal distributed data and for different numbers $M$ of iterations } \label{istonmax_gauss}
\end{center}
\end{figure}

\begin{figure}[!h]
\begin{center}
\includegraphics[height=8cm,width=12cm]{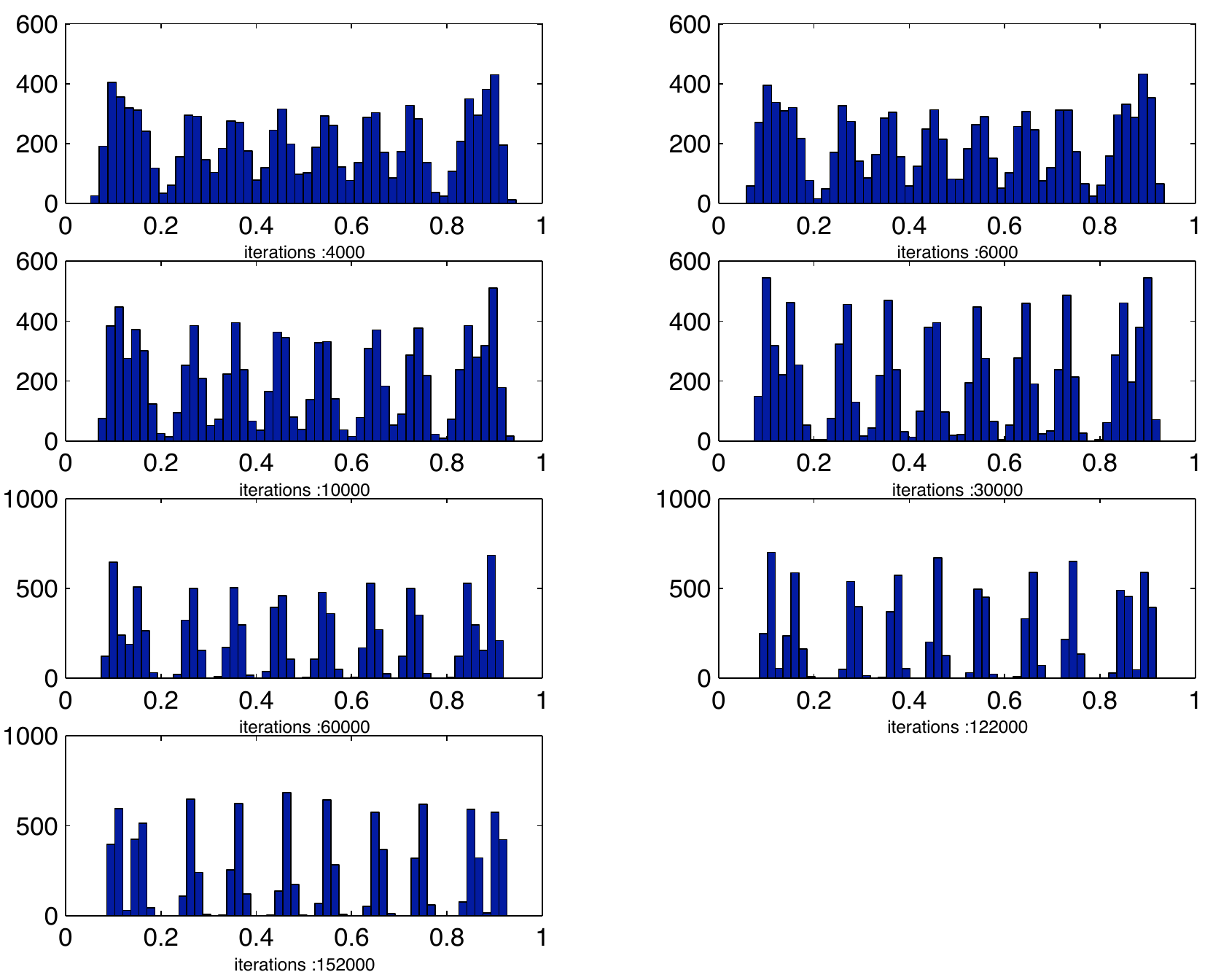}
 \caption{Histograms in the case of $\eta(n) = \frac{\sqrt{6*log(
 n)}}{\sqrt(n)+1}$ and
  uniformly distributed data and for different numbers $M$ of
  iterations obtained repeating the data several times.} \label{istolog_rad1}
\end{center}
\end{figure}

Before explaining our application of Kohonen algorithm to
microarrays data, we make some remarks on the repetitions of data
set presented to the network. This procedure is necessary because
the data set is small for microarray data. The accuracy improves
by increasing the number of samples and this technique does not
change the limit if there is the almost everywhere
convergence.\\To be sure we have done the same analysis
 shown above with a data set of 2000 elements repeated at the beginning 2
 times, then 5, 10,15,30,60 and 125 so to have the same iterations
 of the previous analysis. The results are similar, that is we
 have a.e. convergence for the previous case of $\eta(n)$, that is :
 \begin{itemize}
    \item $\eta(n) = \eta_i(1-\frac{n}{nmax})$
    \item $\eta(n) = \frac{\sqrt{6*log( n)}}{\sqrt(n)+1}$
\end{itemize}
We show the histograms \ref{istonmax1}, \ref{istolog_rad1} in
these two case to illustrate this statement:
\begin{figure}[!h]
\begin{center}
\includegraphics[height=8cm,width=12cm]{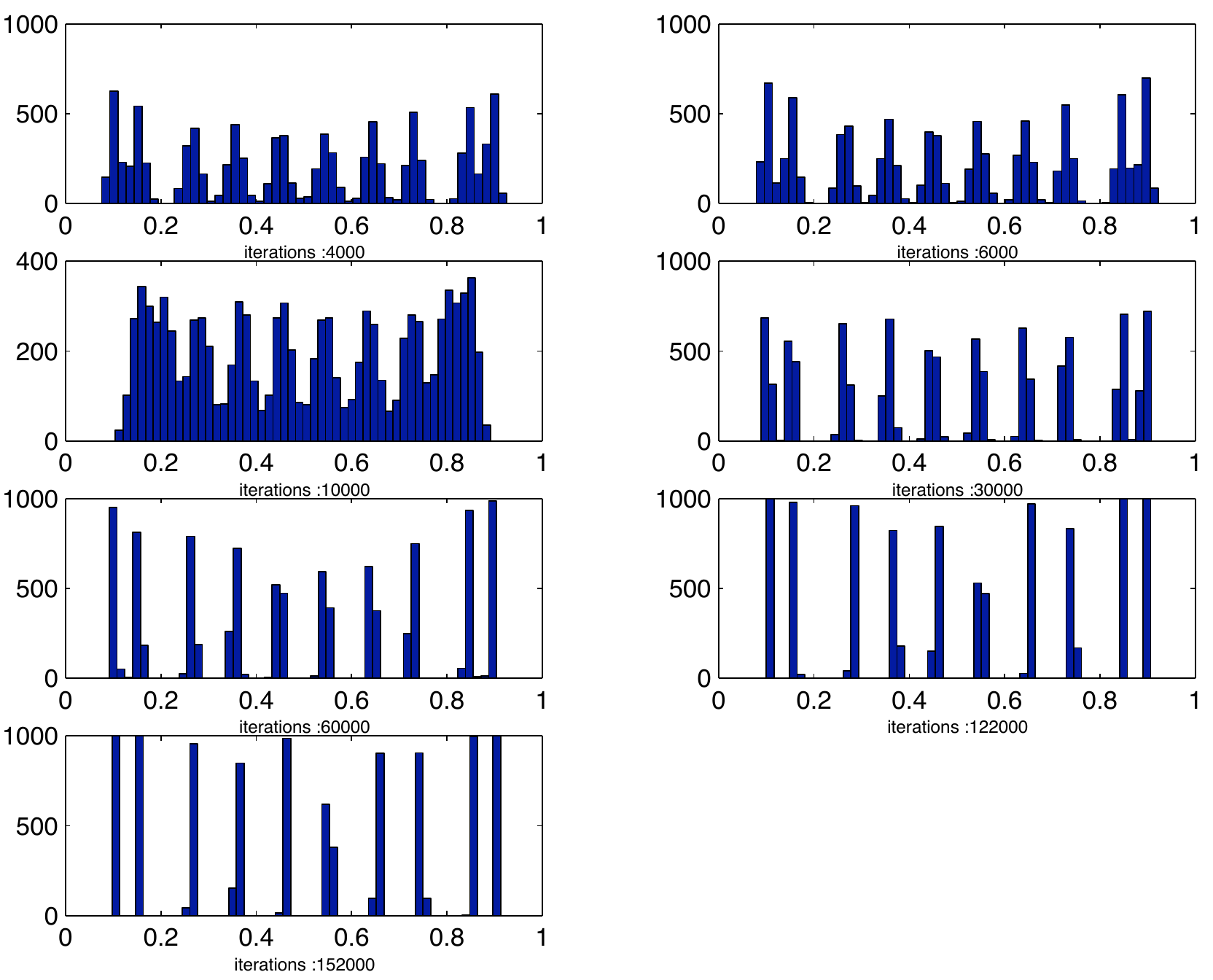}
 \caption{Histograms in the case of $\eta(n) =
 \eta_i(1-\frac{n}{nmax})$ and
  uniformly distributed data and for different numbers $M$ of iterations
 obtained repeating the data several times.} \label{istonmax1}
\end{center}
\end{figure}

\section{ Application to microarrays data}

In this section we show how we have applied the Kohonen network to
micro-arrays data set. Following our strategy we have made the
cluster analysis of the data for each sample and compared the
genes appearing in the nearby clusters, in this way we exploit the
neat convergence properties of the one-dimensional case. The set
of data we analyzed is the same as the one published in
\cite{[30]} where there is an exhaustive description of
microarrays sample preparation. In brief total RNA (ttlRNA) was
extracted and purified from mammary glands in control and
transgenic mice. ttlRNA were pooled to obtain three replicates for
the mammary glands of 2-week-pregnant WT BALB/c mice (wk2prg), of
22 week old untreated BALB-neuT mice(wk22nt), and of 22 week old
primed and boosted BALB-neuT mice(wk22pb)and two replicates for
the mammary glands of 10 week old untreated BALB-neuT mice (
wk10nt). Chips were
scanned to generate digitized image data (DAT) files.\\
DAT files were analyzed by MAS 5.0 to generate background-
normalized image data (CEL files). Probe set intensities were
obtained by means of the robust multi array analysis method
(\cite{[4]}). The full data set was normalized according to the
invariant set method. The full-shaped procedure described by
Saviozzi et al (\cite{[29]}) was then applied. The resulting
$5482$ probe sets were analyzed by combining two statistical
approaches implemented in significance analysis of micro-arrays
(\cite{[3]}): two classes unpaired sample method and the multi
classes response test. This analysis produced a total of 2179
probe sets differentially expressed in at least one of the three
experimental groups. The 2179 probe sets were converted in virtual
two dye experiments comparing all replicates of each experimental
groups with index $j=1,..3$ ( i.e $\frac{wk10nt_i}{wk2prg_j}$
$i=1,2$; $\frac{wk22nt_i}{wk2prg_j}$;$\frac{wk22pb_i}{wk2prg_j}$
$i=1,..3$ ). Therefore we have 3 replicates of 8 experimental groups. \\
 We apply the Kohonen algorithm to the 2179 probe sets.
 Our  main aim is to detect which genes are up-modulated in
 wk22pb respect to wk22nt and wk10nt.\\
The first step is to implement the one dimensional Kohonen
algorithm in Matlab and study  its convergence putting inside as
inputs data the $2179$  expression levels of genes of any
experimental group. In particular our set contains the log values
of expression levels of genes, which are normally distributed in
any experimental group; so,with regard the numerical studies done,
we choose $\eta_i(1-\frac{n}{nmax})$, with $\eta_i=0.8$ because we
have the almost everywhere convergence, and $\Lambda$ as
neighborhood function, since we have the best accuracy in this
case.\\For each experimental group the input set $\Omega$ is only
of $2179$ elements so to improve the accuracy we repeat the data
presentation set several times in different order. We present the
data set $100$ times, in such way the input set is almost $220000$
patterns; in this case the mean error of weights is about $0.01$
(as we have seen in our previous studies) and since the average
variability of the expression levels of genes among the
replicates is about $0.133$, this error is acceptable.\\
We run the Kohonen algorithm fixing $N$, the number of weights,
equal to $30$ and then we choose among the limit values found only
those weights with a distance greater than two times the average
variability of the expression levels of genes among the
replicates. We select the weights in this way because otherwise
the assignment of a gene to a particular cluster could be not
unique. The choice of $N=30$ has been done analyzing the
distribution of the data and considering the variability of the
expression levels of genes among the replicates. To obtain weights
with a distance greater than two times the average variability of
genes expressions we can fix also $N=15$, but in this way we lose
precision in finding weights at the boundaries of the data set
interval. It happens because the data are normally distributed,
therefore they are concentrated near the mean of the data set and
more we move away from mean more the distance between weights
increases, therefore it is better choosing more weights than those
which have an optimal distance between them, such that it is
possible to detect more weights at boundaries, since we want to
find out up-modulated genes.\\Once we have found the limit weights
values we separate the data into the identified clusters. This
procedure has been done for every experimental group indexed by
$j$ ($\frac{wk10nt_i}{wk2prg_j}$ $i=1,2$;
$\frac{wk22nt_i}{wk2prg_i}$;$\frac{wk22pb_i}{wk2prg_i}$
$i=1,..3$), so we have eight classifications for each replicate.
In addition we choose one of the 24 (8 for each replicate)
sequences of limit weights and we separate the data of every
experimental into the clusters identified by the chosen sequence.
In this way we obtained 24 classifications for every sequence of
limit weights
(that are 24).\\
Once we have obtained these classifications we improve the
precision of assignment of genes considering their biological
variability; therefore we have checked if the expression level of
genes which lay on the boundaries of a cluster can be considered
really belonging to that cluster or, because its variability, to
its neighbor. In detail, if the expression of the genes,
incremented of its biological error, is closer to the weight of
its cluster than to its nearest weight, the assignment of the gene
does not change, otherwise the gene is assigned to the cluster
corresponding to its nearest weight.\\
We can observe that, since the limit weights are ordered, the
clusters, with which they are associated  can be ordered in
ascending way. Therefore in clusters related to high index we find
genes
 with a greater expression level than in clusters with low index.\\
For each replicates we select only those genes that are in
clusters with high index for the classifications obtained respect
the limit weights found analyzing the data of
$\frac{wk22pb_i}{wk2prg_i}$ $i=1,..3$ and in low clusters for
classifications obtained by means of the limit weights found
analyzing the data of $\frac{wk10nt_j}{wk2prg_i}$;
$\frac{wk22nt_i}{wk2prg_i}$, $j=1,2$. After this procedure we
 have identified a set of 70 up-modulated genes in wk22pb respect
to wk22nt and wk10nt. Among these genes there are 25 ones that
have not been found by Quaglino et al. These new genes found are
shown in the table \ref {tabella7}.
\begin{table}[!h]
\begin{center}\small
\begin{tabular}{|c|c|c|}
 \hline\\
 Affymetrix ID& Gene Title & Gene Symbol\\
 \hline\\
 100376\_f\_at&similar to immunoglobulin heavy chain &  LOC432710    \\
 101720\_f\_at&immunoglobulin kappa chain variable 8 (V8)&  Igk-V8\\
 101743\_f\_at&immunoglobulin heavy chain 1a (serum IgG2a)& Igh-1a\\
 101751\_f\_at&gene model 194, (NCBI)&Gm194\\
             & gene model 189, (NCBI) & Gm189 \\
             & gene model 192, (NCBI)& Gm192\\
             &  gene model 1068, (NCBI)&Gm1068 \\
             & gene model 1069, (NCBI)&Gm1069\\
             & gene model 1418,(NCBI)&Gm1418\\
             & gene model 1419, (NCBI)&Gm1419\\
             & gene model 1499, (NCBI)&Gm1499\\
             & gene model 1502, (NCBI)&Gm1502\\
             & gene model 1524, (NCBI)&Gm1524\\
             & gene model 1530, (NCBI)&Gm1530\\
             &similar to immunoglobulin light &LOC434586\\
             & chain variable region&LOC545848\\
             &similar to immunoglobulin light chain variable region immunoglobulin light&ad4\\
              &chain variable region gene model 1420, (NCBI)&Gm1420\\
 102722\_g\_at&expressed sequence AI324046&AI324046 \\
 103990\_at&FBJ osteosarcoma oncogene B&Fosb\\
 104638\_at&ADP-ribosyltransferase 1    Art1\\
160927\_at&angiotensin I converting enzyme (peptidyl-dipeptidase A) 1&Ace\\
 161650\_at&secretory leukocyte peptidase inhibitor&Slpi\\
  162286\_r\_at&    Fc fragment of IgG binding protein & Fcgbp\\
 92737\_at&interferon regulatory factor 4 & Irf4\\
 92858\_at&secretory leukocyte peptidase inhibitor& Slpi\\
 93527\_at&Kruppel-like factor 9 &  Klf9\\
  94442\_s\_at&G-protein signalling modulator 3 (AGS3-like, C. elegans)&Gpsm3\\
 94725\_f\_at&similar to immunoglobulin light chain variable region&LOC434033 \\
 96144\_at&    inhibitor of DNA binding 4 & Id4\\
 96963\_s\_at&immunoglobulin light chain variable region&8-30\\
96975\_at&immunoglobulin kappa chain&Igk-V1\\
          & variable 1 (V1)&IgM\\
          &Ig kappa chain&Igk-V5\\
           &immunoglobulin kappa chain&bl1\\
           & variable 5 (V5family)&\\
          & immunoglobulin light chain variable &\\
          &region&\\
 97402\_at&indolethylamine N-methyltransferase&Inmt\\
  97826\_at&Fc fragment of IgG binding protein&  Fcgbp\\
   98452\_at&FMS-like tyrosine kinase 1&  Flt1\\
    98765\_f\_at&similar to immunoglobulin heavy&LOC382653\\
                & chain&LOC544903\\
                &similar to immunoglobulin mu-chain&\\
 99850\_at &Immunoglobulin epsilon heavy chain constant region&\\
102156\_f\_at&immunoglobulin kappa chain&\\
             &constant region &\\
             &mmunoglobulin kappa chain&\\
             & variable 21 (V21)&\\
             & immunoglobulin kappa chain&\\
             & similar to anti-glycoprotein-B of&\\
              &human Cytomegalovirus immunoglobulin Vl chain&\\
               &immunoglobulin kappa chain&\\
               &variable 8 (V8)&\\
              & similar to anti-PRSV coat protein&\\
              &monoclonal antibody PRSV-L 3-8 &\\
              &immunoglobulin light chain variable&\\
              & region&\\
   98475\_at &matrilin 2 & Matn2\\
 \hline
\end{tabular}
\caption{The up modulated genes found out  }\label{tabella7}
\end{center}
\end{table}

\section{Conclusion }

We have improved the theorem on the a.e. convergence of the
Kohonen algorithm because we prove the sufficiency of a slow decay
of the learning parameter, $\sum \eta(n)= \infty$, similar to the
one used in the applications. The theorem is not complete because
we are not able to prove the necessity of such condition and
future work should be concentrated on this point. But for doing
such a research one has to find something functional more similar
to the Liapunov functional than the one currently available. This
could make it possible using some argument of convergence similar
to the one used for the simulated annealing. We made also many
numerical simulations of convergence in order to find the choice
of $\eta(n)$ which minimizes the rate of decrease of the average
error and also for finding which version of the learning algorithm
is better to use. We found that the optimal choice is:

\[\frac{\sqrt{6*log( n)}}{\sqrt(n)+1} \le \eta(n) \le \eta\_i(1-\frac{n}{nmax})\]

The algorithm with $\Lambda$ neighborhood function is better than
the one using the function $h$ since it has bad convergence
properties. The latter one is commonly used in the simulations.
After this detailed analysis we applied our optimal choice to the
genetic expression levels of tumor cells. The 25 genes identified
by us were also consistent with the biological events investigated
by Quaglino (\cite{[30]}).

\end{document}